\documentclass[12pt]{article}
\newcommand{\blind}{0}

\usepackage{booktabs, subcaption, multirow} % Top and bottom rules for table
\usepackage{authblk}
\usepackage{url}
\usepackage{hyperref}
\usepackage{array}
\usepackage{natbib} % Required to change bibliography style to APA
\usepackage{amsthm,amsfonts}
\usepackage{amsmath,amssymb} 
\usepackage{color}
\usepackage{bm}
\usepackage{graphicx}
\usepackage{float} % para [H]
\usepackage{caption}
\usepackage{subcaption}
\usepackage{tabularx}
\usepackage{algorithm}
\usepackage{algpseudocode}
\usepackage{booktabs}
\usepackage{multirow}
\usepackage[most]{tcolorbox} % for colored boxes

\newcommand{\reva}[1]{{\color{red} #1}} % red. 
\newcommand{\revb}[1]{{\color{blue} #1}} % blue

\newcommand{\F}{\textbf{F}}
\newcommand{\C}{\textbf{C}}
\definecolor{lightgrayblue}{RGB}{235, 242, 250}  % subtle blue-gray
\definecolor{darkgrayblue}{RGB}{70, 90, 110}     % for frame
\definecolor{lightbeige}{RGB}{252, 248, 237}     % soft neutral
\definecolor{darkbeige}{RGB}{120, 100, 80}       % for frame
\definecolor{lightgraygreen}{RGB}{235, 250, 240}  % subtle light green
\definecolor{darkgraygreen}{RGB}{60, 120, 80}     % deeper green for frame

\newtcolorbox[auto counter]{promptBox}[2][]{%
  colback=lightgrayblue,
  colframe=darkgrayblue,
  coltext=black,
  fonttitle=\bfseries,
  title=Prompt~\thetcbcounter: #2,
  sharp corners=south,  % optional: modern flat bottom
  boxrule=0.75pt,       % thinner border for elegance
  enhanced
}

\newtcolorbox[auto counter]{promptTemplate}[2][]{%
  colback=lightgrayblue,
  colframe=darkgrayblue,
  coltext=black,
  title=#2,
  fonttitle=\bfseries,
  sharp corners=south,  % optional: modern flat bottom
  boxrule=0.75pt,       % thinner border for elegance
  enhanced,
  #1
}

\newtcolorbox[auto counter]{promptElement}[2][]{%
  colback=lightgraygreen,
  colframe=darkgraygreen,
  coltext=black,
  fonttitle=\bfseries,
  title=#2,
  sharp corners=south,  % optional: modern flat bottom
  boxrule=0.75pt,       % thinner border for elegance
  enhanced,
  #1
}

\newtcolorbox[auto counter]{outputBox}[2][]{%
  colback=lightbeige,
  colframe=darkbeige,
  coltext=black,
  fonttitle=\bfseries,
  title=ChatGPT's Output~\thetcbcounter: #2,
  sharp corners=south,  % optional: modern flat bottom
  boxrule=0.75pt,       % thinner border for elegance
  enhanced
}

\newtcolorbox[auto counter]{outputBoxGemini}[2][]{%
  colback=lightbeige,
  colframe=darkbeige,
  coltext=black,
  fonttitle=\bfseries,
  title=Gemini's Output~A\thetcbcounter: #2,
  sharp corners=south,  % optional: modern flat bottom
  boxrule=0.75pt,       % thinner border for elegance
  enhanced
}

\begin{document}
\def\spacingset#1{\renewcommand{\baselinestretch}%
{#1}\small\normalsize} \spacingset{1}

\title{\bf Prompt engineering using order-of-addition experiments: An application to generating two-level fractional factorial designs}

\if0\blind
{
\author[1]{Duoduo Danny Ying}
\author[2]{Alan R. Vazquez\thanks{Corresponding author, Email: alanrvazquez@tec.mx; ORCID: \texttt{0000-0002-3658-0911}}}
\author[1]{Hongquan Xu\thanks{ORCID: \texttt{0000-0002-3502-0664}}}
\affil[1]{Department of Statistics and Data Science, University of California, Los Angeles, USA}
\affil[2]{School of Engineering and Sciences, Tecnologico de Monterrey, Mexico}
%\affil[$\star$]{Corresponding author: {alanrvazquez@tec.mx}}

 \maketitle
} \fi

\if1\blind
    \author{(Authors)}
     \maketitle
\fi

\begin{abstract}
%The text of your abstract.  100 or fewer words.
\noindent Large language models (LLMs) are becoming ubiquitous in engineering and science because they can turn prompts into data analysis code, experimental designs, formulations of optimization problems, among other applications. However, many LLMs suffer from a phenomenon called order dependency, in which the order of phrases in the prompt affects their performance on a given task. To overcome this issue, we introduce a systematic method that uses order-of-addition experiments to quantify the ordering effect of elements in a prompt and identify their best positions. We demonstrate our methodology by constructing two-level fractional factorial designs using state-of-the-art LLMs. We show that order-of-addition experiments can elucidate order dependency in these LLMs, and can help us to identify a high-quality prompt configuration for the task. 
\end{abstract} 

\noindent%

\spacingset{1.4} %Recommended for explanation document: 1.4 
\pagebreak
%%%%%%%%%%%%%%%%%%%%%%%%%%%%%%%%%%%%%%%%%%%%%%%%%%%%%%%%%%%%%%%%
\section{Introduction} \label{sec:introduction}

In recent years, large language models (LLMs) have helped us tackle a wide variety of tasks due to their remarkable abilities to understand, analyze, and generate text \citep{alammar2024hands}. For instance, they have been used to formulate optimization problems \citep{bertsimas2024robust}, generate code for data analyses \citep{song2026statllm}, build smart reference systems for statistical quality control \citep{megahed2024introducing}, and plan experiments \citep{vazquez2026}. Technically, LLMs are developed from deep neural networks \citep{goodfellow2016deep} that are trained on vast amounts of text data obtained from books, blogs, social media, news and research articles, among other sources. In general, they can be classified in terms of their deep neural network architecture, the predominant one of which is the transformer architecture \citep{vaswani2017attention}. For example, this architecture is in pretrained LLMs implemented in popular generative artificial intelligence chatbots, such as ChatGPT \citep{chatgpt2025} and Gemini \citep{comanici2025gemini}. We refer to \cite{ji2026overview} for a comprehensive treatment of LLMs, including their architectures, training, tuning, and other features.

The main input of an LLM is a prompt, which is a text describing a task to be performed. Ideally, the prompt provides all the necessary information for the LLM to complete the task. To this end, it can contain several elements, such as the role of the LLM, the goal and context of the task, solved examples (called \textit{shots}), and requirements on the output format. The process of writing the prompt that optimizes the performance of the LLM on a given task is known as prompt engineering \citep{schulhoff2024prompt}. 

One of the main challenges in prompt engineering is \textit{order dependency} \citep{mcilroy2024order}, which occurs when the performance of an LLM on a task depends on the order of sentences or phrases within the prompt, despite the semantics remaining identical. \cite{lu2022fantastically}, \cite{pezeshkpour2024large}, and \cite{guan2025order} show that many pretrained transformer-based LLMs exhibit order dependency on a wide variety of tasks and prompts. To overcome this issue, there are several prompt engineering methods in the literature. They include the algorithms of \cite{kumar2021reordering}, \cite{lu2022fantastically}, and \cite{wu2024prompt}, which search for the order of shots in a prompt that optimizes the performance of LLMs on sentiment classification, fact retrieval, textual entailment, and instruction-induction tasks. \cite{pezeshkpour2024large} introduce an algorithm that deals with the ordering of the answers in a multiple-choice question posed in a prompt. Their algorithm runs 10 versions of the prompt, each with a random order of the answers, on the LLM and outputs the answer that was most frequently selected. However, a limitation of these approaches is that they focus only on the ordering of shots or answers. 

Another prompt engineering method to cope with order dependency is the Gradient-free Instructional Prompt Search (GRISP) algorithm of \cite{prasad2023grips}, which does not focus solely on permuting shots or answers in a prompt. Instead, it modifies the phrases in a (predefined) base prompt to find its best formulation for the LLM. Specifically, the GRISP algorithm applies one of four possible changes at random: edit, add, and remove a phrase, and swap two phrases. \cite{prasad2023grips} show that their GRISP algorithm can formulate prompts (with or without shots) that enhance the performance of LLMs on several instruction-based tasks. However, due to its lack of systematic changes, the algorithm may require a large number of executions of the LLMs, which can become expensive when using state-of-the-art paid LLMs, such as those in ChatGPT and Gemini. For the same reason, it also does not allow us to quantify the impact of the changes on the success rate of an LLM on a task.

In this paper, we introduce a new prompt engineering method based on order-of-addition (OofA) experiments that can quantify and model order dependency. OofA experiments study the effect of the order of $q$ components on a response \citep{lin2025order}. Typically, they consist of three stages. First, we conduct an OofA experiment according to an OofA design, which is a highly structured set of component orderings. The OofA design may have far fewer than $q!$ test orderings if $q$ is large \citep{peng2019design,stokes2024metaheuristic}. Next, we analyze the data collected from the experiment using a regression model that links the response to the ordering effects of the components \citep{voelkel2019design,yang2021component,stokes2022position}. Finally, we use the model to identify the order of the components that optimizes the response. In our setup, the components are specific elements in a base prompt, and the response is the success rate of the LLM in achieving a given task. 

Using OofA experiments for prompt engineering provides two main benefits over the aforementioned methods that deal with order dependency. First, through the use of regression models, we can estimate the effect of the order of elements in a prompt (e.g., sentences, phrases, shots, and answers) on the success rate of an LLM. To our knowledge, the estimation of ordering effects in prompts is new to the prompt engineering literature. Second, through the use of cost-efficient OofA designs, we can limit the number of executions of the LLM needed to optimize a prompt. This computational efficiency is particularly valuable when the number of prompt components, $q$, is large. 

We demonstrate OofA experiments for prompt engineering by analyzing and optimizing a base prompt that instructs an LLM to construct two-level fractional factorial designs \citep{vazquez2026}. The prompt has several elements including the role of the LLM, the instruction, the output format, the Zero-shot-CoT technique \citep{kojima2022large}, and an example of an optimal two-level design. We use these elements because they have been shown to improve the performance of pretrained transformer-based LLMs across a variety of tasks \citep{schulhoff2024prompt}. In our application, we focus on tasks that require constructing 16-run two-level designs with seven to 12 factors, and primarily use \textbf{gpt-4.1} \citep{openai2024gpt41} and \textbf{gemini-2.5-flash} \citep{comanici2025gemini} as our LLMs. We do this because these tasks are known to be challenging for these LLMs \citep{vazquez2026}. 

Our application involves three OofA experiments to study the effect of modifying the elements of our base prompt. In the first experiment, we study permutations and rephrased versions of the instruction in the base prompt for tasks with seven to 12 factors using \textbf{gpt-4.1}. We show that the formulation of the instruction element has an effect on the performance of this LLM, and that this effect is different across tasks. In the second experiment, we permute the order of all elements in the base prompt to construct a 16-run 9-factor design using \textbf{gpt-4.1}. We show that OofA designs and models allow us to identify high-quality orderings of the prompt elements. In particular, we show how to increase the success rate in constructing the 9-factor 16-run optimal design from 11.67\% under the base prompt to 98.3\% under the optimized prompt configuration. In the third experiment, we apply OofA experiments to  \textbf{gemini-2.5-flash} and show that we can identify a prompt configuration that increases the success rate of constructing a 16-run 7-factor optimal design from 35\% to 100\%. Collectively, these experiments show that our prompt engineering method based on OofA experiments can estimate the ordering effects of prompt elements, efficiently identify high performing prompt element orderings, and be applied to different tasks and LLMs.

The remainder of the paper is organized as follows. In Section~\ref{sec:OofA_models}, we introduce the design and analysis of OofA experiments. In Section~\ref{sec:promptsystem}, we present the base prompt and the LLMs used in our three experiments. In Section~\ref{sec:results}, we discuss the design and analysis of the first two experiments and, to limit the length of the paper, defer the third experiment to Supplementary Section~S3. In Section~\ref{sec:conclusion}, we end the article with concluding remarks that include comments on the latest versions of the GPT and Gemini models at the time of writing the article. The supplementary materials include sections with supporting information for the main manuscript and R and Python programs to reproduce our study.

%%%%%%%%%%%%%%%%%%%%%%%%%%%%%%%%%%%%%%%%%%%%%%%%%%%%%%%%%%%%%%%%
\section{Design and modeling for order-of-addition experiments}\label{sec:OofA_models}

OofA experiments study the order in which $q$ components are used in a system. We denote the $q$ components as $1, \ldots, q$. Following \cite{prasad2023grips}, we use phrases and sentences as components, as well as different levels of a component corresponding to different ways of rephrasing it. Our setup allows us to maintain the general structure of the prompt, while providing enough flexibility for its optimization. We now introduce the statistical models and designs used in our OofA experiments for prompt engineering.  

\subsection{Logistic regression model} \label{sec:logistic_models}

If there are few components, an initial model to capture the potential relationship between their positions in a prompt and the success rate of an LLM in a task is the logistic regression model. This is because, in prompt engineering, the response under study ($Y$) is typically the number of times the LLM successfully completed a task in $N_r$ independent repetitions. So, $Y \sim \text{Bin}(N_r, p)$, where $p$ is the probability of success (or success rate) in the task. 

In this initial model, the $q!$ orderings of the $q$ components are seen as the levels of a categorical variable, which is included in the model using dummy variables as follows. Let $O$ be the set of all $q!$ orderings in lexicographical order. For example, for $q=3$, $O = \{123, 132, 213, 231, 312, 321\}$. Let $s_i$ be a binary dummy variable that takes the value of one if and only if the $i$-th element in $O$ was used to obtain $Y$, where $ i = 1, \ldots, q!$. The initial logistic regression model is
\begin{equation}\label{eq:logisticmodel}
    (Y|\mathbf{s}) \sim \text{Bin}(N_r, p(\mathbf{s})) \text{ with } p(\mathbf{s})^{-1} = 1+ e^{-(\alpha_0 + \sum_{i=1}^{q!-1} \alpha_i s_{i+1})},
\end{equation}
\noindent where $s_1$ is the \textit{reference} ordering and $\mathbf{s} = (s_2, \ldots, s_{q!})^T$. For $i \geq 1$, the value of $\alpha_i$ is the change in the probability of success given by the $i$-th ordering, relative to that of the reference ordering. Excluding $\alpha_0$, if one of these $\alpha_i$'s  differs from zero, the LLM exhibits order dependency. %% I think \reva{($i>0$)} is redundant since we say if one of these referring to the  

Following \citet{tsai2023dual}, we extend the initial logistic model to accommodate two levels of each of the $q$ components. Here, the two levels of a component are given by the original phrase (or sentence) and a predefined rephrase version. For example, a component in one of our OofA experiments to construct two-level fractional factorial designs is the phrase ``$n$ runs and $m$ factors''. One level of this component is given by the original phrase, and the other by the simple rephrase ``$m$ factors and $n$ runs''. 

We use the variable $w_i$ to denote the version of the $i$-th component. Specifically, $w_i$ equals $-1$ or $1$ if and only if the component is in its original or rephrased form, respectively. Moreover, we collect the versions of all components in the vector $\mathbf{w}=(w_1, \ldots, w_q)^T$. In this model, $(Y|\mathbf{s}, \mathbf{w}) \sim \text{Bin}(N_r, p(\mathbf{s}, \mathbf{w}))$ and 
\begin{equation}\label{eq:extendedlogisticmodel}
    p(\mathbf{s}, \mathbf{w})^{-1} = 1+ e^{-(\alpha_0 + \sum_{i=1}^{q!-1} \alpha_i s_{i+1} + \sum_{i=1}^{q} \gamma_i w_i)},
\end{equation}
\noindent where $\gamma_i$ is the coefficient associated with the two versions of component $i$. A negative value of $\gamma_i$ implies that the original version of component $i$ increases the probability of success in the task, while a positive value implies that this version reduces this probability. If $\gamma_i = 0$, the probability of success does not depend on the version of the $i$-th component.

We use maximum likelihood estimation to estimate the logistic regression models in Equations~\eqref{eq:logisticmodel} and \eqref{eq:extendedlogisticmodel}, and use Wald tests to identify significant coefficients. We evaluate the goodness of fit of a model using McFadden's pseudo $R^2$ \citep{menard2000coefficients}, which is one minus the log-likelihood ratio between the full and null models. The higher the value of this statistic, the better the fit of the model to the data. Additionally, we quantify overdispersion using the parameter 
\begin{equation}
    \hat{\phi} = \frac{\sum_{i=1}^{N_t} {\hat{r}^{2}_i}}{N_t-p},
    \label{eq:disp}
\end{equation}
where $\hat{r}_i$ is the Pearson residual of the model, $p$ is the number of terms in it (including the intercept), and $N_t$ is the total number of observations. A value of $\hat{\phi}$ greater than one indicates the presence of overdispersion, meaning that the observed variability in the response exceeds the expected variability under the assumed binomial model. If this is the case, we adjust the standard errors of the coefficient estimates by a factor of $\sqrt{\hat{\phi}}$, and recompute the p-values of the Wald tests with this adjustment.

\subsection{Logistic pairwise-ordering model} \label{sec:logisticPWOmodel}

The logistic regression models in the previous section are useful to identify general differences between the orders and versions of the components on the success rate of an LLM. However, they do not allow us to study  individual effects of placing one component before another. To overcome this issue, \citet{van1995design} introduced the pairwise-ordering (PWO) model, which is defined using pseudo factors that represent the relative ordering of two components. A pseudo factor between components $i$ and $j$ is 
\begin{equation*}
    z_{ij} = \begin{cases} \phantom{-}1 & \text{if component } i \text{ precedes component } j, \\
    -1 & \text{otherwise.}
    \end{cases}
\end{equation*}
\noindent To study $q$ components, we need $q(q-1)/2$ pseudo factors. 

The PWO model was originally developed for modeling quantitative responses with or without multiple levels of the components; see \cite{van1995design} and \cite{tsai2023dual}. Here, we extend it to model our binomial response $Y$. Let $\mathbf{z} = (z_{12}, z_{13}, \ldots, z_{(q-1)q})^T$. Our model is  
\begin{equation}\label{eq:compoundPWOmodel}
    (Y|\mathbf{z}, \mathbf{w}) \sim \text{Bin}(N_r, p(\mathbf{z}, \mathbf{w})) \text{ with }     p(\mathbf{z},\mathbf{w})^{-1} = 1 + e^{-(\beta_0 + \sum_{i=1}^{q-1}\sum_{j=i+1}^{q} \beta_{ij} z_{ij} + \sum_{i=1}^q \gamma_i w_i)},
\end{equation}
where $\beta_0$ is the intercept and $\beta_{ij}$ is the coefficient associated with ordering component $i$ before $j$. In particular, a positive value of $\beta_{ij}$ implies that placing component $i$ before $j$ increases the probability of success in the task. A negative value implies the opposite, that placing component $i$ after $j$ increases this probability. If $\beta_{ij}=0$, the probability of success is not affected by the order between components $i$ and $j$. The interpretation of the $w_i$'s and $\gamma_i$'s and the estimation of the model in Equation~\eqref{eq:compoundPWOmodel} are  as in Section~\ref{sec:logistic_models}. Following \cite{tsai2023dual}, we call this model the \textit{compound} logistic PWO model, which, to our knowledge, is new to the literature on OofA experiments.

In some experiments, such as the ones in Section~\ref{sec:GPT41results} and Supplementary Section S3, %\ref{sec:Geminiresults}, 
restricting to the main effects of the $z_{ij}$'s and $w_{i}$'s may not be enough to model the probability of success of the LLM. For these cases, we consider a \textit{compound interaction} logistic PWO model that includes two-way interactions among these terms. In this model, the distribution of $Y$ is the same as before, but the exponent of $p(\mathbf{z},\mathbf{w})$ in Equation~\eqref{eq:compoundPWOmodel} has additional interaction terms that are divided into three groups. The first group contains $q^2(q-1)^2/8 - q(q-1)/4$ interaction terms involving two different $z_{ij}$'s. The second group has $q^2(q-1)/2$ interaction terms that involve a $z_{ij}$ and a $w_i$. The last group has $q(q-1)/2$ interaction terms involving two different $w_i$'s.

The total number of terms in the full PWO interaction model is  $(q^2-q+2)(q^2+3q+4)/8$, 
which quickly increases with the number of components $q$. % \revb{(The same as ${q(q-1)/2+q \choose 2} + {q(q-1)/2+q \choose 1} + 1$.)} 
For situations in which the model is too large, we can use step-wise regression guided by information criteria (such as the Bayesian information criterion) to identify a good submodel. We refer to \cite{claeskens2008model} for details on model selection and information criteria for logistic regression. We also refer to \cite{mee2020order} for a comprehensive treatment of interactions among the pseudo factors in the PWO model for a continuous response.

\subsection{Full and fractional designs} \label{sec:fullfactorialdesign}

In the absence of component levels, we can conduct an OofA experiment using the full ordering design with $q!$ tests when the value of $q$ is small. For example, if $q$ is three or four, this design has six and 24 tests, respectively. If each component has two levels, we may use the full ordering factorial (OF) design with $(q!)(2^q)$ tests \citep{yang2023ordering} provided that $q$ is small. For $q=3$, the number of tests is 48, which was feasible in our first OofA experiment in Section~\ref{sec:GPTresults}. However, as the number of components gets larger than three or four, the number of tests in these designs quickly becomes too expensive.

To overcome this issue, we can use a fractional OofA design given by a well-chosen subset of tests from the full design. For the PWO model of \cite{van1995design} and the compound PWO model of \cite{tsai2023dual}, there are several fractional designs with fewer tests than $q!$ or $(q!)(2^q)$; see \cite{voelkel2019design}, \cite{peng2019design}, \cite{chen2020construction}, \cite{zhao2021designs}, \cite{schoen2023order} and \cite{Zhao02102025}, for example. However, in addition to the fact that they assume a continuous response, these designs are available for specific numbers of tests, which prevents them from being flexible.  

In the literature on optimal experimental designs, there are two main approaches to constructing designs for logistic regression models \citep{berger2009introduction}. One approach necessitates good initial values for the coefficients in Equation~\eqref{eq:compoundPWOmodel}, and the other assumes that the coefficients are random variables that follow a user-specified prior distribution. However, in prompt engineering, it is difficult to specify these values or prior distribution due to the lack of practical applications of our logistic PWO models. 

To overcome this issue, we use a simple approach to constructing designs for our second and third OofA experiments in Section~\ref{sec:results} and Supplementary Section S3. Specifically, we use the algorithm of \cite{stokes2022position}, which constructs fractional OofA designs with $q$ components, flexible numbers of tests, and high estimation efficiency for the standard PWO model. Our choice for this algorithm is also motivated by the fact that, when the run size exceeds the number of pseudo factors, the resulting design resembles a component orthogonal array \citep[COA;][]{yang2021component}, which is a highly structured fractional OofA design. In fact, when the number of tests, $N$, is a multiple of $q(q-1)$, the design obtained by the algorithm is guaranteed to be a COA. In this design, each component appears equally often at each position. Moreover, for any two positions, each component combination $(i,j)$ with $i \neq j$ also appears equally often. \cite{yang2021component} and \cite{stokes2022position} show that COAs are robust under different OofA models beyond PWO-based models.   

Algorithm~\ref{alg:coa} shows the pseudo code for the algorithm of \cite{stokes2022position}, which works as follows. The inputs to the algorithm are the desired number of tests ($N$) and the number of components ($q$), which must be a prime power. In the algorithm, the $q$ components are labeled as $0, \ldots, q-1$, and the operations are performed over a Galois field of order $q$. That is, $GF(q) = \{\omega_0,\omega_1,\dots,\omega_{q-1}\}$, where $\omega_0$ is the zero element \citep{barker1986sum}. The algorithm starts by creating $q-1$ basic matrices $\mathbf{L}_k$ whose entries are obtained by adding and multiplying elements of $GF(q)$. Next, a parent matrix $\mathbf{C}_1$ is constructed by vertical concatenation of all basic matrices $\mathbf{L}_k$. After that, the algorithm generates $(q-2)!$ child matrices from the parent matrix $\mathbf{C}_1$. The first of these matrices is $\mathbf{C}_1$ and the rest are obtained by permuting the last $q-2$ columns of $\mathbf{C}_1$. We denote these child matrices as $\C_2, \dots, \C_{(m-2)!}$. The algorithm continues by constructing a generator matrix $\mathbf{F}_q$ by row-wise concatenation of $\C_1, \C_2, \dots, \C_{(q-2)!}$. The basic OofA design $\mathbf{F}_{N,q}$ is obtained from the first $N$ rows of $\mathbf{F}_q$. Finally, the basic design $\mathbf{F}_{N,q}$ is improved by permuting its columns to optimize, for instance, the estimation efficiency of the PWO model, measured by the $D$ criterion \citep{atkinson2007optimum}. The improved design $\mathbf{F}_{N,q}$ is the output of the algorithm. 

If the $q$ components have two levels, we can construct a fractional OF design using Kronecker products involving $\mathbf{F}_{N,q}$ and a two-level full or fractional factorial design \citep{tsai2023dual,yang2023ordering}. Specifically, let $\mathbf{X}$ be the two-level $N_f$-run design involving the different versions of the $q$ components. The fractional OF design is $\left[ \mathbf{F}_{N,q} \otimes \mathbf{1}_{N_f}, \mathbf{1}_N \otimes \mathbf{X} \right]$, where $\mathbf{1}_{N}$ is the $N \times 1$ column of ones.

\begin{algorithm}[ht]
\caption{Pseudocode of the algorithm of \cite{stokes2022position}.} \label{alg:coa}
\begin{algorithmic}[1]
% <-- this [1] turns on line numbering (Step 1, 2, ...)
\renewcommand{\algorithmicrequire}{\textbf{Input:}}
\renewcommand{\algorithmicensure}{\textbf{Output:}}
\Require $N$ and prime power $q$.
\State For $k = 1, \dots, q-1$, define a $q\times q$ matrix $\textbf{L}_k = (l^{(k)}_{ij})$ such that $l^{(k)}_{ij} = \omega_i + \omega_k \omega_j$, for $i,j = 0, \dots, q-1$, where the addition and multiplication are defined on $GF(q)$.
\State Construct a $(q^2-q)\times q$ matrix $\C_1$ by row-wise concatenating $\textbf{L}_1,\dots,\textbf{L}_{q-1}$.
\State Keep the first two columns of $\C_1$ fixed and permute the last $q-2$ columns of $\C_1$ in a systematic way. This results in $(q-2)!$ permuted matrices, denoted as $\C_1, \dots, \C_{(q-2)!}$.
\State Construct a $q!\times q$ matrix $\F_q$ by row-wise concatenating $\C_1,\dots,\C_{(q-2)!}$ and replacing $\omega_i$ with number $i$, for $i = 0, \dots, q-1$.
\State Let $\F_{N,q}$ be the $N \times q$ design formed by the first $N$ rows of $\F_q$.
\State Permute the columns of $\F_{N,q}$ to improve its performance under a chosen criterion.
\Ensure $\F_{N,q}$.
\end{algorithmic}
\end{algorithm}

%%%%%%%%%%%%%%%%%%%%%%%%%%%%%%%%%%%%%%%%%%%%%%%%%%%%%%%%%%%%%%%%

\section{Constructing two-level fractional factorial designs using an LLM} \label{sec:promptsystem}

To demonstrate the effectiveness of OofA experiments for prompt engineering, we consider the construction of two-level fractional factorial designs using LLMs as \cite{vazquez2026}. These authors showed that Generative Pre-trained Transformer (GPT) models and Gemini models can construct good designs with up to 16 runs and nine factors. In our application, we use a base prompt, \textbf{gpt-4.1} and \textbf{gemini-2.5-flash} as the LLMs, and statistical criteria to evaluate the resulting design. We discuss each of these building blocks separately, starting with the statistical criteria because it is used in our base prompt.

\subsection{Criteria for two-level fractional factorial designs}

Two-level fractional factorial designs are classified as regular and nonregular \citep{wu2011experiments}. Regular designs have a defining relation and a run size that must be a power of two. In contrast, nonregular designs do not have a defining relation, but their run sizes are a multiple of four. Following \cite{vazquez2026}, we focus on regular two-level designs because they are commonly taught in experimental design courses \citep{vazquez2025review} and are widely used in practice \citep{mee2009comprehensive}. Because of their dominance over nonregular designs, these designs are simply called two-level fractional factorial designs. 

To evaluate a two-level design (regular or nonregular), we use the generalized word length pattern \citep[GWLP;][]{xu2001generalized}. Let $m$ and $n$ be the numbers of factors and runs in the two-level design, respectively. Let $\mathbf{X}_k$ be an $n \times p$ matrix containing the contrast vectors of the $k$-factor interactions, where $p = {m \choose k}$. Let $A_k = n^{-2}\| \mathbf{1}^{T}_n \mathbf{X}_k\|^2_{2}$, where $\mathbf{1}_n$ is the $n \times 1$ vector of ones and $\| \mathbf{x} \|_{2}$ is the $L_2$-norm of vector $\mathbf{x}$. The GWLP of a two-level design is $(A_1, A_2, \ldots, A_m)$. The generalized minimum aberration criterion \citep{xu2001generalized} is to sequentially minimize the GWLP from left to right. The design that achieves this is called the generalized minimum aberration design which, among all comparable designs, has the smallest aliasing between low-order effects (e.g., main effects and two-factor interactions). 

If the design is regular, the GWLP reduces to the wordlength pattern containing the frequencies of the lengths of words in its defining relation. In this case, the generalized minimum aberration criterion reduces to the minimum aberration criterion and the resolution of a design is the smallest index $i$ for which $A_i >0$ \citep{wu2011experiments}. A minimum aberration design has maximum resolution, but the reverse is not necessarily true. Throughout the paper, we refer to the minimum aberration design as the optimal design for simplicity.

\subsection{Base prompt} \label{sec:promptelements}

Our base prompt for constructing two-level fractional factorial designs using an LLM has five prompt elements. The first four are taken from \cite{vazquez2026}. Specifically, the first element informs the LLM that it must be an expert in the field of design of experiments when constructing a design. It is based on the prompting technique called role prompting \citep{wang2024role}, which assigns a role to the LLM to focus its expertise on a given domain. We call this element ROLE and show it below. 

\begin{promptElement}{ROLE} 
You are an expert in the subfield of statistics called design of experiments.
\end{promptElement}

The second element taken from \cite{vazquez2026} informs the LLM about its goal, which is to construct the best two-level fractional factorial in terms of resolution and minimum aberration. This element also specifies the number of runs ($n$), the number of factors ($m$), and the coded levels of the design. We call this element INSTRUCTION.

\begin{promptElement}{INSTRUCTION} 
Your goal is to construct a two-level fractional factorial design with maximum resolution and minimum aberration. The number of factors is $m$ and the number of runs is $n$. The factors have two levels coded as `-1' and `+1'.
\end{promptElement}

The third element involves the prompting technique called Zero-shot-CoT, which has been shown to improve the performance of pretrained transformer-based LLMs on a variety of tasks \citep{kojima2022large}. The technique allows the LLM to break a task into smaller intermediate steps, solve each one, and thereby complete the overall task more effectively. Following \cite{vazquez2026}, this element requests that the LLM think step by step when constructing the design. For this reason, we call this element STEP\_BY\_STEP below. 

\begin{promptElement}{STEP\_BY\_STEP}
You will think step by step about how to construct the design.
\end{promptElement}

The fourth element is called FORMAT. It informs the LLM that the first row and column of the design table must contain the labels of the factors and runs, respectively. It also informs the LLM that the full design table must be in comma-separated values (CSV). This is needed because we use the Python functions from \cite{vazquez2026} to evaluate a two-level design, which requires it to be in this format. Another feature of FORMAT is that it instructs the LLM to produce a design table only, avoiding the explanation of the design and its properties. This element is shown below.

\begin{promptElement}{FORMAT} 
You will only generate a table containing the design. You will not generate any text explaining the table or your answer. The table must be in a comma-separated values (CSV) format. Specifically, the values in the table must be separated by ‘,’ and each row must end with ‘\textbackslash\textbackslash’. In the table, the first row will be used as a header row to label the factors using the letters in the English alphabet starting with ‘A’. The first column will be called “Run” and used to count the number of runs starting at ‘1’. Each design cell (excluding the header and Run columns) must contain either ‘-1’ or ‘1’.
\end{promptElement}

The last element of the base prompt is a shot involving an optimal design in terms of resolution and minimum aberration. \cite{brown2020language} shows that adding shots to a prompt enhances the performance of pretrained transformer-based LLMs in question answer, translation, and completion tasks. Despite this fact, \cite{vazquez2026} did not use shots in their prompt to construct two-level fractional factorial designs. They argued that the development and inclusion of shots in a prompt is challenging and may significantly increase the length of the prompt, causing it to be more expensive to run. To overcome this issue, the last element of our base prompt, called SHOT, is a simple example of an optimal design with eight runs and four factors. 

\begin{promptElement}{SHOT} 
As an example, here is an 8-run 4-factor design with maximum resolution and minimum aberration: Run,A,B,C,D \textbackslash\textbackslash  1,-1,-1,-1,-1 \textbackslash\textbackslash 2,-1,-1,1,1 \textbackslash\textbackslash 3,-1,1,-1,1 \textbackslash\textbackslash 4,-1,1,1,-1 \textbackslash\textbackslash 5,1,-1,-1,1 \textbackslash\textbackslash 6,1,-1,1,-1 \textbackslash\textbackslash 7,1,1,-1,-1 \textbackslash\textbackslash 8,1,1,1,1.
\end{promptElement}

\noindent The motivation behind SHOT is threefold. First, \cite{vazquez2026} show that GPT and Gemini LLMs have a 100\% success rate in generating the optimal 8-run 4-factor design in terms of resolution and minimum aberration. Second, the example design is shown in CSV format, which further encourages the LLM to follow the output format in FORMAT. Third, Supplementary Section S1 shows  the significant effect of this element on generating the optimal two-level design with 16 and nine factors. To our knowledge, the use of SHOT in a prompt to construct two-level fractional factorial designs is new to the literature. 

Using the five elements described previously, we develop Prompt \ref{ptp:basic} below, which is our base prompt to construct a two-level fractional factorial design.

\begin{promptTemplate}[label ={ptp:basic}]{Prompt 1} 
\{ROLE\} \{INSTRUCTION\} \{STEP\_BY\_STEP\} \{FORMAT\} \{SHOT\}
\end{promptTemplate}

\subsection{LLMs and their setup} \label{sec:setupLLM} 

In our application, we use \textbf{gpt-4.1} and \textbf{gemini-2.5-flash} because they have demonstrated performance in tasks that involve coding, long input contexts, and following instructions \citep{openai2024gpt41api,comanici2025gemini}. We use these LLMs through the Python packages called \verb!openai! \citep{openai_python} and \verb!google-genai! \citep{google-genai}, and API keys from OpenAI and Google. However, the use of these LLMs has associated costs in terms of their input and output tokens. Essentially, tokens are the data units of an LLM and are constructed from text using a process called \textit{tokenization}, which considers language punctuation, and context, among other features \citep{alammar2024hands}. OpenAI charges \$2 and \$8 for one million tokens involved in input and output text, respectively, of \textbf{gpt-4.1} \citep{openai2024gpt41}. For \textbf{gemini-2.5-flash}, these costs of input and output text from Google are \$0.3 and \$2.5, respectively \citep{google_gemini_api_pricing_2025}. Since prompt optimization often requires hundreds or even thousands of LLM executions, the cumulative token cost can become substantial. Therefore, it is important to use cost-efficient OofA designs for prompt engineering with these LLMs.

Both \textbf{gpt-4.1} and \textbf{gemini-2.5-flash} have tuning parameters that drive their performance on a task. One of the most important ones is \textit{temperature} because it controls the diversity and randomness of the model's output \citep{alammar2024hands}. A higher value of this parameter generally results in a more diverse output, while a lower value creates a more deterministic output. In our application, we set \textit{temperature} to zero to test the basic knowledge of \textbf{gpt-4.1} and \textbf{gemini-2.5-flash}, and limit the noise in the data collected from them. This enhances the reproducibility of our OofA experiments conducted with these LLMs.

Another important tuning parameter of \textbf{gemini-2.5-flash} is \textit{thinkingBudget}, which defines the strategy for reasoning of this LLM. Because our goal is to test the basic knowledge of this LLM, we set this parameter to zero, which turns the reasoning feature off. In any case, in Section~\ref{sec:conclusion}, we comment on  recent LLMs that have a reasoning feature and show that they also exhibit order dependency in some of our tasks.

%%%%%%%%%%%%%%%%%%%%%%%%%%%%%%%%%%%%%%%%%%%%%%%%%%%%%%%%%%%%%%%%

\section{Application} \label{sec:results}

Our application of OofA designs and models for prompt engineering involves three experiments. In the first experiment, we demonstrate the presence of order dependency in \textbf{gpt-4.1} by modifying the INSTRUCTION element of Prompt~\ref{ptp:basic}. In the second experiment, we show how OofA experiments allow us to efficiently identify orders of the elements in Prompt~\ref{ptp:basic} that maximize the success rate of \textbf{gpt-4.1}. Finally, the third experiment demonstrates the generalization of our approach by applying OofA experiments to \textbf{gemini-2.5-flash}. To save space, we discuss the third experiment in  Supplementary Section~S3.

\subsection{First experiment with gpt-4.1}\label{sec:GPTresults}

In this experiment, we investigate the effect of phrase ordering within INSTRUCTION. Specifically, we consider a modified version of it called INSTRUCTION(X-Y-Z).
\begin{promptElement}{INSTRUCTION(X-Y-Z)} 
Your goal is to construct a two-level fractional factorial design with \{X\}. The design has \{Y\}. The design has \{Z\}.
\end{promptElement}

\noindent The \{X\}, \{Y\}, and \{Z\} positions can each contain one of the following three phrases: ``maximum resolution and minimum aberration,'' ``$m$ factors and $n$ runs,'' and ``two levels for each factor coded as $-1$ and $+1$.'' These phrases define the three components in the first experiment. We refer to these components as criteria, design dimensions, and level coding, and label them as 1, 2, and 3, respectively. For each component, we consider two levels (versions) coded as $-1$ and 1. The low  level corresponds to the original phrase, while the high level corresponds to a semantically equivalent phrase. Specifically, the equivalent phrases for components 1, 2 and 3 are ``minimum aberration and maximum resolution,'' ``$n$ runs and $m$ factors,'' and ``two levels for each factor coded as $+1$ and $-1$'', respectively. 

The goal of the experiment is to detect differences in the success rates of \textbf{gpt-4.1} for constructing 16-run designs with seven to 12 factors, caused by the different component orderings and versions in INSTRUCTION(X-Y-Z). In what follows, we first show the OofA design used for collecting data in this experiment. Next, we analyze the data using descriptive statistics and the logistic regression model in Section~\ref{sec:logistic_models}. 

\subsubsection{Data collection}

We use a full OF design with $3! \times 2^3 = 48$ tests \citep{yang2023ordering}. We denote each test in this design using the actual sequence of the components and their levels used, given by the variables $w_1$, $w_2$, and $w_3$. For example, the test with the sequence 231 and $w_1=-1$, $w_2=1$, and $w_3=-1$, implies that the design dimensions, level coding, and criteria come in the \{X\}, \{Y\}, and \{Z\} position, respectively. Moreover, design dimensions is being used in its rephrased version, while the other components are in their original version. The final instruction for this test is shown below.

\begin{promptElement}{INSTRUCTION(2-3-1)} 
Your goal is to construct a two-level fractional factorial design with $n$ runs and $m$ factors. The design has two levels for each factor coded as $-1$ and $+1$. The design has maximum resolution and minimum aberration.
\end{promptElement}

For each test in the full OF design, we insert the resulting instruction into the prompt template shown in Prompt~\ref{ptp:study1} below. We then submit Prompt~\ref{ptp:study1} to \textbf{gpt-4.1} with the number of runs and factors in the two-level design construction task. 

\begin{promptTemplate}[label ={ptp:study1}]{Prompt 2} 
\{ROLE\} \{INSTRUCTION(X-Y-Z)\} \{STEP\_BY\_STEP\} \{FORMAT\} \{SHOT\}
\end{promptTemplate}

For each task, we conducted the full OF design with 40 replicates in a completely randomized way. Table \ref{tb:3parts_seq} shows the success rates of \textbf{gpt-4.1} given by the different tests for the six 16-run design construction tasks with seven to 12 factor. The first three columns of the table show the levels of the three components, and the following 12 columns show the success rates (or probabilities) of the LLM for the different component orderings and levels. The component orderings used to obtain each success rate are shown in bold. 

\begin{table}[]
    \caption{Success rates of \textbf{gpt-4.1} for constructing 16-run optimal designs with seven to 12 factors obtained from the first OofA experiment.}\label{tb:3parts_seq}
    \renewcommand{\tabcolsep}{4pt}
    \centering
    \small
    \begin{tabular}{rrr|llllll|llllll}
\toprule
& &  \multicolumn{13}{c}{Number of Factors} \\ \cline{4-15}

            $w_1$ & $w_2$ & $w_3$ & 7          & 8         & 9         & 10        & 11        & 12 & 7          & 8         & 9         & 10        & 11        & 12 \\
\hline
 & & & \multicolumn{6}{c|}{\textbf{123}} & \multicolumn{6}{c}{\textbf{231}} \\
 \hline
 $-1$ & $-1$ & $-1$ & 0.250 & 1.000 & 0.150 & 0.900 & 0.175 & 0.825 & 0.000 & 0.750 & 0.400 & 0.600 & 0.225 & 0.750 \\
 $-1$ & $-1$ & 1 & 0.100 & 0.875 & 0.100 & 0.875 & 0.175 & 0.800 & 0.350 & 0.975 & 0.325 & 0.725 & 0.075 & 0.725\\
 $-1$ & 1 & $-1$ & 0.150 & 0.825 & 0.050 & 0.725 & 0.250 & 0.725 & 0.400 & 0.500 & 0.200 & 0.625 & 0.300 & 0.600\\
 $-1$ & 1 & 1 & 0.700 & 0.425 & 0.000 & 0.675 & 0.325 & 0.800 & 0.525 & 0.125 & 0.075 & 0.925 & 0.275 & 0.625\\
 1 & $-1$ & $-1$ & 0.625 & 0.950 & 0.125 & 0.925 & 0.150 & 0.825 & 0.125 & 0.325 & 0.550 & 0.600 & 0.200 & 0.675\\
 1 & $-1$ & 1 & 0.525 & 0.900 & 0.125 & 0.950 & 0.125 & 0.875 & 0.425 & 0.950 & 0.225 & 0.725 & 0.125 & 0.800\\
 1 & 1 & $-1$ & 0.850 & 0.800 & 0.100 & 0.750 & 0.175 & 0.825 & 0.475 & 0.100 & 0.400 & 0.650 & 0.075 & 0.875\\
 1 & 1 & 1 & 0.550 & 0.925 & 0.025 & 0.850 & 0.275 & 0.750 & 0.125 & 0.375 & 0.175 & 0.825 & 0.125 & 0.650\\[1ex]
 \hline
 & & & \multicolumn{6}{c|}{\textbf{132}} & \multicolumn{6}{c}{\textbf{312}} \\
 \hline
 $-1$ & $-1$ & $-1$ & 0.050 & 0.675 & 0.125 & 0.975 & 0.100 & 0.875 & 0.075 & 0.175 & 0.125 & 0.900 & 0.050 & 0.825\\
 $-1$ & $-1$ & 1 & 0.325 & 0.625 & 0.200 & 0.950 & 0.175 & 0.950 & 0.375 & 0.000 & 0.025 & 0.925 & 0.075 & 0.825\\
 $-1$ & 1 & $-1$ & 0.150 & 0.100 & 0.025 & 0.800 & 0.025 & 0.500 & 0.350 & 0.200 & 0.075 & 0.500 & 0.125 & 0.200\\
 $-1$ & 1 & 1 & 0.150 & 0.025 & 0.000 & 0.950 & 0.250 & 0.700 & 0.500 & 0.000 & 0.100 & 0.775 & 0.075 & 0.275\\
 1 & $-1$ & $-1$ & 0.100 & 0.475 & 0.175 & 0.950 & 0.200 & 0.925 & 0.425 & 0.875 & 0.150 & 0.975 & 0.050 & 0.900\\
 1 & $-1$ & 1 & 0.225 & 0.350 & 0.250 & 0.975 & 0.025 & 1.000 & 0.725 & 0.350 & 0.125 & 0.825 & 0.075 & 0.950\\
 1 & 1 & $-1$ & 0.300 & 0.200 & 0.100 & 0.700 & 0.475 & 0.850 & 0.325 & 0.825 & 0.100 & 0.625 & 0.075 & 0.550\\
 1 & 1 & 1 & 0.050 & 0.125 & 0.025 & 0.775 & 0.175 & 0.800 & 0.250 & 0.250 & 0.100 & 0.850 & 0.175 & 0.650\\[1ex]
 \hline
 & & & \multicolumn{6}{c|}{\textbf{213}} & \multicolumn{6}{c}{\textbf{321}} \\
 \hline
$-1$ & $-1$ & $-1$ & 0.275 & 0.400 & 0.100 & 0.750 & 0.225 & 0.500 & 0.150 & 0.650 & 0.325 & 0.700 & 0.225 & 0.725\\
$-1$ & $-1$ & 1 & 0.200 & 0.950 & 0.025 & 0.525 & 0.025 & 0.500 & 0.400 & 0.700 & 0.225 & 0.700 & 0.175 & 0.725\\
$-1$ & 1 & $-1$ & 0.400 & 0.925 & 0.000 & 0.200 & 0.400 & 0.675 & 0.125 & 0.575 & 0.125 & 0.550 & 0.050 & 0.225\\ 
$-1$ & 1 & 1 & 0.700 & 0.750 & 0.025 & 0.575 & 0.050 & 0.450 & 0.125 & 0.825 & 0.100 & 0.450 & 0.025 & 0.050\\
1 & $-1$ & $-1$ & 0.400 & 0.775 & 0.175 & 0.725 & 0.075 & 0.425 & 0.225 & 0.875 & 0.250 & 0.850 & 0.150 & 0.875\\
1 & $-1$ & 1 & 0.525 & 0.900 & 0.100 & 0.550 & 0.050 & 0.675 & 0.450 & 0.900 & 0.200 & 0.950 & 0.025 & 0.900\\
1 & 1 & $-1$ & 0.500 & 0.925 & 0.050 & 0.475 & 0.100 & 0.400 & 0.000 & 0.375 & 0.125 & 0.300 & 0.000 & 0.325\\
1 & 1 & 1 & 0.575 & 0.450 & 0.075 & 0.700 & 0.075 & 0.475 & 0.050 & 0.625 & 0.150 & 0.575 & 0.125 & 0.100\\
\bottomrule
\end{tabular}
\end{table}

\subsubsection{Data analysis}

\begin{figure}
    \centering
    \includegraphics[width=\textwidth]{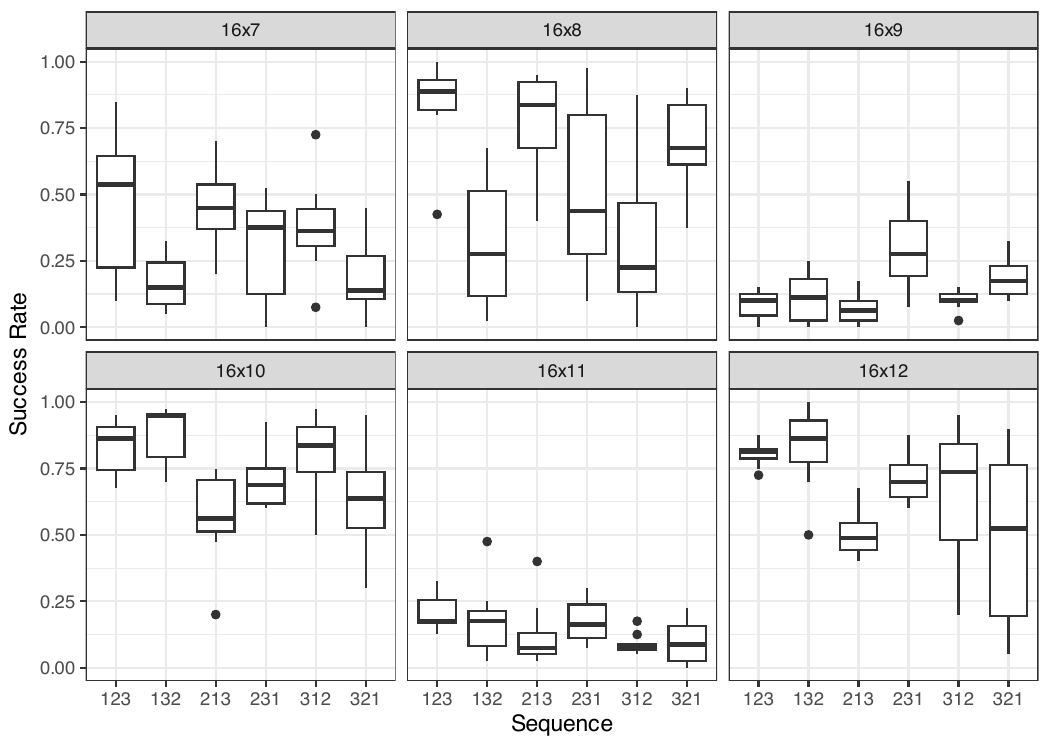}
    \caption{Success rates of \textbf{gpt-4.1} for constructing 16-run optimal designs with seven to 12 factors across sequences.}
    \label{fig:seq_boxplot}
\end{figure}

We conduct an exploratory analysis of the data in Table~\ref{tb:3parts_seq}. Figure \ref{fig:seq_boxplot} shows boxplots of the distributions of the success rates of the LLM for each combination of component orderings and tasks. There are clear differences among the orderings within a task. For example, when the task has $m = 8$ factors, the sequences $123$ and $132$ produce substantially different success rates. This indicates that, with criteria fixed at the beginning, placing design dimensions before level coding improves the success rate for this task. When $m = 12$, the sequences $123$ and $213$ also exhibit differences, indicating that, with level coding fixed at the end, placing criteria before design dimensions tends to improve the success rate.

\begin{figure}
    \centering
    \includegraphics[width=\textwidth]{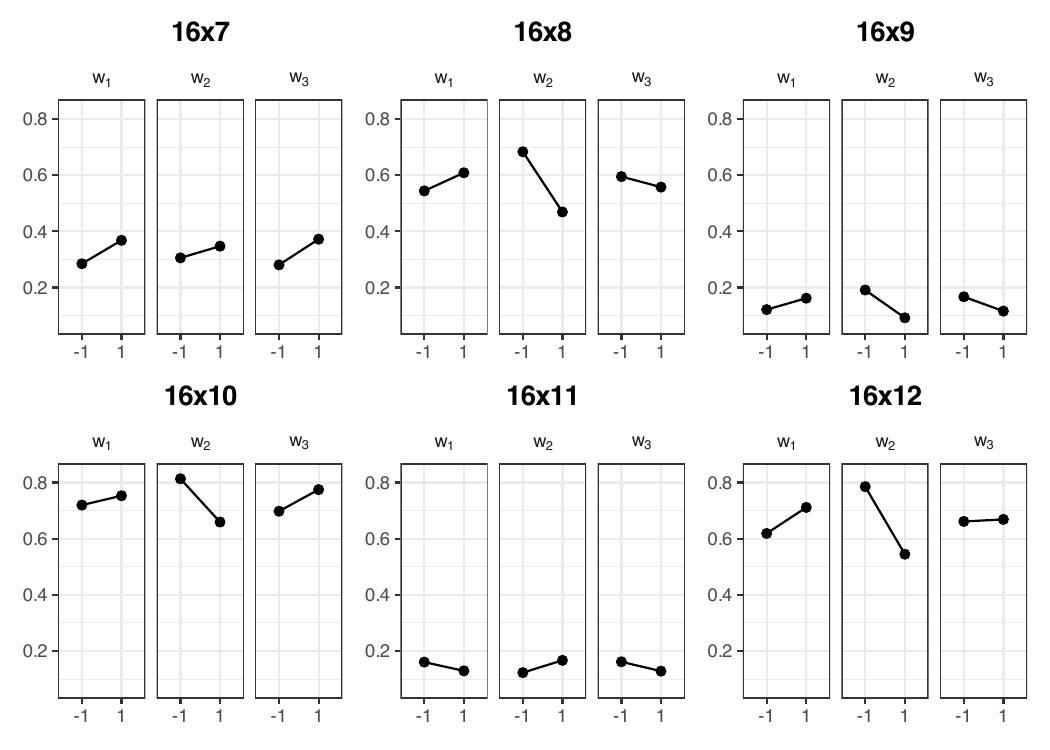}
    \caption{Success rates of \textbf{gpt-4.1} for constructing 16-run optimal designs with seven to 12 factors across versions.}
    \label{fig:version_boxplot}
\end{figure}

In addition to the effect of component ordering, we examine the effect of their versions across tasks. Figure~\ref{fig:version_boxplot} shows main effect plots for $w_1$, $w_2$, and $w_3$ for the six design construction tasks. For tasks with eight, nine, 10, and 12 factors, the version of design dimensions ($w_2$) appears to have the strongest main effect on the success rate. Specifically, the high level of this component---which specifies the number of runs before the number of factors---is associated with a lower success rate than the low level that involves the original phrase. Overall, Figures \ref{fig:seq_boxplot} and  \ref{fig:version_boxplot} suggest that both prompt component orderings and versions influence the success rate of \textbf{gpt-4.1} in generating optimal two-level designs.

To quantify the effects of component orderings and versions, we fit the logistic regression model in Equation~\eqref{eq:extendedlogisticmodel} with $q=3$ to the data in Table~\ref{tb:3parts_seq}. Table~\ref{tb:3parts} shows the estimated coefficients and some statistics of the model for each design construction task. To further assess the relative importance of the orderings and versions, Table~S2 in the supplementary sections provides the sequential analysis of deviance for the logistic regression model, along with the percentage of explained deviance attributable to each term.

The results confirm that both the orderings and versions of the prompt components contribute substantially to the variability in performance. Regarding the orderings, Table~\ref{tb:3parts} shows that each of the dummy variables for the orderings is significant for the success rate in at least one task. Regarding the versions, the table shows that the effect of $w_2$ is significant at a level of $\alpha=0.01$ in at least four tasks. For these cases, the estimated value of $\gamma_2$ is negative, indicating that the rephrased version of design dimensions decreases the success rate of the LLM in generating optimal designs with eight, nine, 10, and 12 factors. In other words, it is more effective to write ``$m$ factors and $n$ runs'' in Prompt~\ref{ptp:study1}. Consistent with this finding, the effect of $w_2$ accounts for between 12\% and 28\% of the explained deviance in four of our six tasks; see Table~S2. This result highlights the sensitivity of \textbf{gpt-4.1} performance to seemingly minor changes in prompt component rephrasing.

\iffalse %%
\begin{table}%[h!]
\centering
\caption{Estimated coefficients and associated statistics.}\label{tb:3parts}
\begin{tabular}{lllllll}
\toprule
            & 7          & 8         & 9         & 10        & 11        & 12 \\ \midrule
$\hat{\beta_0}$ & -0.775 *** & 0.355 .   & -2.001 ***& 1.156 *** & -1.826 ***& 0.791 *** \\
$\hat{\beta}_{12}$    & 0.253      & -0.333    & -0.162    & 0.606 *** & 0.147     & 0.565 *** \\
$\hat{\beta}_{13}$    & -0.078     & 0.284     & -0.474 ***& -0.085    & 0.084     & -0.027   \\
$\hat{\beta}_{23}$    & 0.493 **   & 0.374     & 0.17      & 0.031     & 0.218     & 0.258 . \\
$\hat{\gamma}_1$       & -0.201     & -0.153    & -0.18 .   & -0.095    & 0.129     & -0.237 . \\
$\hat{\gamma}_2$       & -0.101     & 0.494 *   & 0.442 *** & 0.435 *** & -0.18     & 0.598 *** \\
$\hat{\gamma}_3$       & -0.22 .    & 0.089     & 0.226 *   & -0.219 *  & 0.137     & -0.019   \\
\bottomrule
Null deviance     & 425.94  & 914.34 & 194.42    & 341.36    & 160.79    & 492.24\\
Residual deviance & 312.86  & 641.39 & 75.65     & 154.72    & 127.67    & 243.58\\
\reva{$R^2_{McF}$} & 0.265  & 0.299  & 0.611     & 0.547     & 0.206     & 0.505\\
\reva{$\hat{\phi}$}& 6.763  & 15.036  & 1.886    & 3.58      & 3.032      & 5.613 \\
\bottomrule
Signif. codes: &  0 `***' & 0.001 `**' & 0.01 `*' & 0.05 `.' & 0.1 ` ' & 1 \\
\bottomrule
\end{tabular}
\end{table}
\fi %%

\begin{table}%[h!]
\centering
\caption{Estimated coefficients and associated statistics of a logistic regression model for 16-run construction tasks with 7 to 12 factors.}\label{tb:3parts}
\begin{tabular}{lllllll}
\toprule
&  \multicolumn{6}{c}{Number of Factors} \\ \cline{2-7}
           & 7          & 8         & 9         & 10        & 11        & 12 \\ \midrule
Intercept & $-$0.128 & \phantom{$-$}1.753** & $-$2.502*** & \phantom{$-$}1.677*** & $-$1.368*** & \phantom{$-$}1.537***\\
$s_2$ & $-$1.500** & $-$2.558*** & \phantom{$-$}0.327 & \phantom{$-$}0.453 & $-$0.183 & \phantom{$-$}0.154\\
$s_3$ & $-$0.090 & $-$0.516 & $-$0.226 & $-$1.41*** & $-$0.604 & $-$1.482***\\
$s_4$ & $-$0.724. & $-$1.698** & \phantom{$-$}1.566*** & $-$0.732* & $-$0.205 & $-$0.537\\
$s_5$ & $-$0.382 & $-$2.498*** & \phantom{$-$}0.192 & $-$0.236 & $-$1.005* & $-$0.867*\\
$s_6$ & $-$1.349** & $-$0.885 & \phantom{$-$}0.947*** & $-$1.092*** & $-$0.892* & $-$1.579***\\
$w_1$ & \phantom{$-$}0.206. & \phantom{$-$}0.169 & \phantom{$-$}0.183* & \phantom{$-$}0.096 & $-$0.130 & \phantom{$-$}0.245*\\
$w_2$ & \phantom{$-$}0.103 & $-$0.550** & $-$0.452*** & $-$0.441*** & \phantom{$-$}0.182 & $-$0.618***\\
$w_3$ & \phantom{$-$}0.226. & $-$0.098 & $-$0.231** & \phantom{$-$}0.222* & $-$0.138 & \phantom{$-$}0.019\\ 
\midrule
Null deviance & 425.940 & 914.340 & 194.420 & 341.360 & 160.790 & 492.240\\
Deviance & 263.130 & 459.310 & 46.380 & 130.850 & 113.870 & 184.870\\
$R^2_{McF}$ & 0.382 & 0.498 & 0.761 & 0.617 & 0.292 & 0.624\\
$\hat\phi$ & 6.015 & 11.196 & 1.084 & 3.199 & 2.777 & 4.511\\
\bottomrule
Signif. codes: &  0 `***' & 0.001 `**' & 0.01 `*' & 0.05 `.' & 0.1 ` ' & 1 \\
\bottomrule
\end{tabular}
\end{table}

Overall, these results provide formal statistical support for the patterns observed in Figures \ref{fig:seq_boxplot} and \ref{fig:version_boxplot}. They demonstrate that both prompt component orderings and versions significantly affect the success rate of design construction using \textbf{gpt-4.1}, underscoring the importance of systematic prompt optimization.

\subsection{Second Experiment with gpt-4.1} \label{sec:GPT41results}

The goal of our second OofA experiment is to identify the order of the five elements in Prompt~\ref{ptp:basic} that maximizes the success rate of \textbf{gpt-4.1}. Here, the five elements define the five components, with components 1, 2, 3, 4, and 5 corresponding to ROLE, INSTRUCTION, STEP\_BY\_STEP, FORMAT, and SHOT, respectively, as introduced in Section~\ref{sec:promptelements}. Building upon the results in the previous section, we also consider two levels for component 2. That is, the low level is the original INSTRUCTION element, and the high level is given by the semantically equivalent version below, which we call INSTRUCTION$_1$. 
\begin{promptElement}{INSTRUCTION$_1$} 
Your goal is to construct a two-level fractional factorial design with maximum resolution and minimum aberration. The number of runs is $n$ and the number of factors is $m$. The factors have two levels coded as `-1' and `+1'.
\end{promptElement}
\noindent In INSTRUCTION$_1$, we replace the second sentence in INSTRUCTION with ``The number of runs is $n$ and the number of factors is $m$.'' 

In this experiment, we focus on constructing the 16-run 9-factor two-level optimal design, because \cite{vazquez2026} showed that this task is challenging for Gemini and GPT models. As in the previous section, we first discuss the design used for conducting the experiment and then the results from the data analysis.

\subsubsection{Data collection}

A full OF design has $(5!)(2) = 240$ tests, which becomes computationally expensive when we consider multiple replicates. To overcome this issue, we constructed a 40-run COA design using Algorithm \ref{alg:coa} with $q=5$. Specifically, we first generated a $20\times 5$ matrix $\C_1$. Next, we applied the permutation (1,2,3,5,4) to the columns of $\C_1$ to obtain $\C_2$. After that, we concatenated $\C_1$ and $\C_2$ to generate $\F_{40,5}$. We then applied the permutation (1,4,3,2,5) to $\F_{40,5}$ to improve the $D$-criterion of the standard PWO model. We obtain the final OF design as $\left[ \mathbf{F}_{40,5} \otimes \mathbf{1}_{2}, \mathbf{1}_{40} \otimes \mathbf{w}_2 \right]$, where $\mathbf{w}_2 = (-1, 1)^T$ contains the levels of component 2. This OF design has 80 runs.

Table~\ref{tb:40coadesign} shows the 80-run OF design. To help the reader, we show the test prompt produced by test 41 (with sequence 14325 and $w_2=1$) of this OF design below. 

%14325 
\begin{promptTemplate}[label ={ptp:test}]{Test Prompt} 
\{ROLE\} \{FORMAT\} \{STEP\_BY\_STEP\} \{INSTRUCTION$_1$\}   \{SHOT\}
\end{promptTemplate}
 
\noindent We conducted the 80-run OF design as a completely randomized design replicated 20 times. Compared to the previous experiment, we used a smaller number of replicates due to the larger number of tests of the OF design. For each test, we constructed a test prompt like the one above and submitted it to \textbf{gpt-4.1}. Table~\ref{tb:40coadesign} shows the success rate (among the 20 replicates) of this LLM in constructing the optimal 16-run 9-factor design for each test.

\begin{table}
\centering
\small
\caption{Ordering factorial design and success rates of \textbf{gpt-4.1} for constructing optimal 16-run 9-factor designs.} %\reva{ (Prompt~\ref{ptp:basic} only refers to 12345 ($-1$)?)}
\label{tb:40coadesign}
\begin{tabular}{ccrc|ccrc|ccrc}
\toprule
Test & Sequence & $w_2$ & Rate & Test & Sequence & $w_2$ & Rate & Test & Sequence & $w_2$ & Rate\\ \hline
1 & 14325 & $-1$ & 0.35 & 28 & 31254 & $-1$ & 0.50 & 55 & 54132 & 1 & 0.00\\
2 & 25431 & $-1$ & 0.40 & 29 & 42315 & $-1$ & 0.15 & 56 & 13452 & 1 & 0.40\\
3 & 31542 & $-1$ & 0.90 & 30 & 53421 & $-1$ & 0.55 & 57 & 24513 & 1 & 0.40\\
4 & 42153 & $-1$ & 0.05 & 31 & 13245 & $-1$ & 0.15 & 58 & 35124 & 1 & 0.30\\
5 & 53214 & $-1$ & 0.80 & 32 & 24351 & $-1$ & 0.25 & 59 & 41235 & 1 & 0.15\\
6 & 12534 & $-1$ & 0.35 & 33 & 35412 & $-1$ & 0.55 & 60 & 52341 & 1 & 0.00\\
7 & 23145 & $-1$ & 0.25 & 34 & 41523 & $-1$ & 0.30 & 61 & 15324 & 1 & 0.50\\
8 & 34251 & $-1$ & 0.30 & 35 & 52134 & $-1$ & 1.00 & 62 & 21435 & 1 & 0.20\\
9 & 45312 & $-1$ & 0.20 & 36 & 12453 & $-1$ & 0.15 & 63 & 32541 & 1 & 0.60\\
10 & 51423 & $-1$ & 0.10 & 37 & 23514 & $-1$ & 0.20 & 64 & 43152 & 1 & 0.00\\
11 & 15243 & $-1$ & 0.80 & 38 & 34125 & $-1$ & 0.25 & 65 & 54213 & 1 & 0.10\\
12 & 21354 & $-1$ & 0.30 & 39 & 45231 & $-1$ & 0.00 & 66 & 14532 & 1 & 0.40\\
13 & 32415 & $-1$ & 0.30 & 40 & 51342 & $-1$ & 0.60 & 67 & 25143 & 1 & 0.10\\
14 & 43521 & $-1$ & 0.10 & 41 & 14325 & 1 & 0.45 & 68 & 31254 & 1 & 0.80\\
15 & 54132 & $-1$ & 0.65 & 42 & 25431 & 1 & 0.10 & 69 & 42315 & 1 & 0.20\\
16 & 13452 & $-1$ & 0.60 & 43 & 31542 & 1 & 0.25 & 70 & 53421 & 1 & 0.00\\
17 & 24513 & $-1$ & 0.45 & 44 & 42153 & 1 & 0.00 & 71 & 13245 & 1 & 0.20\\
18 & 35124 & $-1$ & 0.50 & 45 & 53214 & 1 & 0.00 & 72 & 24351 & 1 & 0.25\\
19 & 41235 & $-1$ & 0.30 & 46 & 12534 & 1 & 0.15 & 73 & 35412 & 1 & 0.05\\
20 & 52341 & $-1$ & 0.70 & 47 & 23145 & 1 & 0.25 & 74 & 41523 & 1 & 0.05\\
21 & 15324 & $-1$ & 0.90 & 48 & 34251 & 1 & 0.15 & 75 & 52134 & 1 & 0.55\\
22 & 21435 & $-1$ & 0.00 & 49 & 45312 & 1 & 0.25 & 76 & 12453 & 1 & 0.05\\
23 & 32541 & $-1$ & 0.45 & 50 & 51423 & 1 & 0.00 & 77 & 23514 & 1 & 0.00\\
24 & 43152 & $-1$ & 0.10 & 51 & 15243 & 1 & 0.65 & 78 & 34125 & 1 & 0.35\\
25 & 54213 & $-1$ & 0.50 & 52 & 21354 & 1 & 0.30 & 79 & 45231 & 1 & 0.00\\
26 & 14532 & $-1$ & 0.55 & 53 & 32415 & 1 & 0.35 & 80 & 51342 & 1 & 0.05\\
27 & 25143 & $-1$ & 0.10 & 54 & 43521 & 1 & 0.10 &  &  &  & \\
\bottomrule
\end{tabular}
\end{table}

\subsubsection{Data Analysis}

\begin{figure}[h]
    \centering
    \includegraphics{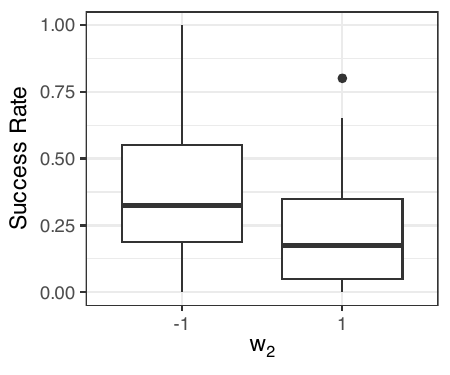}
    \caption{Success rates of \textbf{gpt-4.1} for constructing the 16-run 9-factor optimal design across two versions of INSTRUCTION.}
    \label{fig:FactorRun-GPT}
\end{figure}

Our exploratory analysis of the data in Table~\ref{tb:40coadesign} revealed an interesting result regarding the effect of the two levels of component 2. Specifically, we found substantial differences in success rates for several paired test runs that differ only in the level of component 2, including tests 3 and 43, 5 and 45, 15 and 55, and 20 and 60. To further investigate this effect, Figure~\ref{fig:FactorRun-GPT} presents box plots of the success rates obtained by \textbf{gpt-4.1} under the $-1$ and 1 level of component 2. The figure shows that the success rate tends to be higher when $w_2 = -1$ (corresponding to the original INSTRUCTION) than when $w_2 = 1$ (corresponding to INSTRUCTION$_1$). In other words, specifying the desired number of factors before the desired number of runs tends to have a higher success rate than the reverse ordering. This is consistent with the first experiment.

%%Moreover, when the number of runs was stated before the number of factors, the LLM occasionally generated $32 \times 9$ designs instead of the requested $16 \times 9$ designs. This suggests that the ordering of information within a prompt can influence how the LLM interprets numerical specifications. One possible explanation is that the model places greater emphasis on information appearing later in the instruction. This hypothesis differs from some prior studies on the positional bias in LLMs, which often report a primacy bias whereby information presented near the beginning of the context exerts a stronger influence on LLM \citep{liu2024lost}. Overall, these results indicate that even a minor change in the ordering of prompt components can have a substantial impact on the generated output.

To test  the significance of $w_2$, we fitted a logistic PWO model to the data in Table \ref{tb:40coadesign} with
\begin{equation}
p(\mathbf{z},\mathbf{w})^{-1} = 1 + e^{-(\beta_0 + \sum_{i=1}^{4}\sum_{j=i+1}^{5} \beta_{ij} z_{ij} + \gamma_2 w_2)}.
\label{eq:model_2}
\end{equation}
Table~\ref{tb:40coa} shows the estimated coefficients and a summary of fit of this model. The estimated dispersion parameter $\hat{\phi}$ is higher than one, indicating the presence of overdispersion. We thus multiplied the standard errors for all coefficient estimates by $\sqrt{\hat{\phi}}$ and adjusted the p-values in Table~\ref{tb:40coa} accordingly. We see that the most important variable is $w_{2}$ and it is highly significant because its p-value is less than 0.001. The coefficient of $w_{2}$ is  negative, confirming our conclusion from Figure~\ref{fig:FactorRun-GPT} that placing the number of factors before the number of runs in the prompt increases the success rate in obtaining optimal 16-run 9-factor designs.

\begin{table}%[ht]
\centering
\caption{Estimated coefficients and associated statistics of an initial logistic PWO model for the 16-run 9-factor design task.} \label{tb:40coa}
\begin{tabular}{lrrrl}
\toprule 
            & Estimate  & Std.\ Error & z value & Pr($>|z|$) \\
\midrule 
Intercept & $-$0.951    & 0.125     & $-$7.592    & 3.16e-14  \\
$z_{12}$    & $-$0.039    & 0.179     & $-$0.216    & 0.829  \\
$z_{13}$    &  0.264    & 0.180      &  1.466    & 0.143   \\
$z_{14}$    & $-$0.150     & 0.177     & $-$0.845   & 0.398   \\
$z_{15}$    & 0.291     & 0.181     &  1.604   & 0.109 \\
$z_{23}$    & $-$0.370     & 0.179     & $-$2.060    & 0.039 \\
$z_{24}$    &  0.434    & 0.182     &  2.387   & 0.017 \\
$z_{25}$    & $-$0.334    & 0.178     & $-$1.880    & 0.060  \\
$z_{34}$    & 0.122     & 0.176     &  0.693   & 0.488  \\
$z_{35}$    & 0.057     & 0.179     &  0.316    & 0.752  \\
$z_{45}$    & $-$0.321    & 0.180     & $-$1.786    & 0.074   \\
$w_{2}$   & $-$0.461     & 0.122     & $-$3.772  & 1.62e-4 \\
\midrule 
\multicolumn{5}{l}{Null deviance: 534.82 on 79 degrees of freedom} \\
\multicolumn{5}{l}{Residual deviance: 332.34 on 68 degrees of freedom} \\
$R^2_{McF}$       & 0.379 \\
$\hat\phi$        & 4.266 \\
\bottomrule 
\end{tabular}
\end{table}

Other important variables in Table~\ref{tb:40coa} are $z_{23}$, $z_{24}$, $z_{25}$ and $z_{45}$, with p-values less than 0.1. The estimates of these coefficients indicate that placing component 3 and 5 before component 2, and placing component 4 after component 2, is associated with a higher success rate in generating the optimal 16-run 9-factor design with \textbf{gpt-4.1}. Therefore, according to the model, we recommend a prompt in which STEP-BY-STEP and SHOT precede INSTRUCTION, which, in turn, precedes FORMAT. Moreover, in this prompt, we should place ``number of factors is 9'' before ``number of runs is 16'' in INSTRUCTION.

Table~\ref{tb:40coa} shows that $R^2_{McF}=0.379$ for the model. However, it turns out that it can be improved by including two-way interactions among the pseudo factors and $w_2$. The full interaction model has a total of 66 terms excluding the intercept. To find a reduced model, we applied step-wise regression to the model using the Bayesian information criterion \citep{claeskens2008model}. Table \ref{tb:40coa_model_2} shows the final reduced model  from this approach. Compared to the model in Table~\ref{tb:40coa}, the model in this table has 10 additional interaction terms and $R^2_{McF}=0.773$. The estimated dispersion parameter $\hat{\phi}$ value is also reduced by almost one half compared to that in Table~\ref{tb:40coa}, indicating a substantially improved fit. 

Table~\ref{tb:40coa_model_2} shows that the variables $z_{24}$, $w_{2}$, $z_{14}z_{15}$, $z_{14}z_{25}$, $z_{24}z_{25}$, $z_{13}z_{15}$, and $z_{12}z_{45}$ are highly significant as their corresponding p-values are below 0.001. The p-values and standard errors in Table~\ref{tb:40coa_model_2} were adjusted for overdispersion. In conclusion, two-way interactions of the pseudo factors are important to model the success rate of \textbf{gpt-4.1} in obtaining the optimal 16-run 9-factor design. 

\begin{table}%[ht]
\centering
\caption{Estimated coefficients and associated statistics of a logistic regression model with interaction effects for the 16-run 9-factor construction task.}
\label{tb:40coa_model_2}
\begin{tabular}{lrrrl}
\toprule 
            & Estimate  & Std.\ Error & z value & Pr($>|z|$) \\
\midrule 
(Intercept) & $-$1.099    & 0.123     & $-$8.927    & $<2e-16$  \\
$z_{12}$    & 0.242    & 0.127     & 1.906    & 0.057  \\
$z_{13}$    & 0.232    & 0.133      &  1.750    & 0.08   \\
$z_{14}$    & $-$0.082     & 0.142     & $-$0.575   & 0.565   \\
$z_{15}$    & 0.073     & 0.139     &  0.526   & 0.599 \\
$z_{24}$    & 0.494    & 0.142     &  3.486   & 4.9e-4 \\
$z_{25}$    & $-$0.332    & 0.144     & $-$2.308    & 0.021  \\
$z_{34}$    & 0.181     & 0.125     &  1.448   & 0.148  \\
$z_{35}$    & 0.271     & 0.143     &  1.901    & 0.057  \\
$z_{45}$    & $-$0.150    & 0.132     & $-$1.140    & 0.254  \\
$w_{2}$   & $-$0.540     & 0.098     & $-$5.505  & 3.7e-8  \\
$z_{14}z_{25}$   & $-$0.443     & 0.118     & $-$3.760  & 3.43e-4  \\
$z_{14}z_{15}$   & 0.736     & 0.13     & 5.684  & 1.7e-4  \\
$z_{24}z_{25}$   & $-$0.443     & 0.122     & $-$3.640  & 1.32e-8  \\
$z_{13}z_{15}$   & $-$0.505     & 0.125     & $-$4.046  & 2.73e-4  \\
$z_{12}z_{45}$   & 0.277     & 0.106     & 2.627  & 5.21e-5  \\
$z_{34}z_{45}$   & $-$0.319     & 0.116     & $-$2.752  & 0.009  \\
$z_{13}z_{25}$   & $-$0.295     & 0.118     & $-$2.506  & 0.006  \\
$z_{25}w_{2}$   & 0.385     & 0.108     & 3.581  & 0.012  \\
$z_{35}w_{2}$   & 0.328     & 0.113     & 2.894  & 0.004  \\
$z_{34}w_{2}$   & $-$0.295     & 0.106     & $-$2.785  & 0.005  \\
$z_{14}w_{2}$   & 0.180     & 0.100     & 1.797  & 0.072 \\
\midrule 
\multicolumn{5}{l}{Null deviance: 534.82 on 79 degrees of freedom}\\
\multicolumn{5}{l}{Residual deviance: 121.66 on 58 degrees of freedom}
 \\
$R^2_{McF}$       & 0.773 \\
$\hat\phi$        & 2.1 \\
\bottomrule 
\end{tabular}
\end{table}

Of course, a limitation of the model in Table~\ref{tb:40coa_model_2} compared to that in Table~\ref{tb:40coa} is its interpretability. However, the model in Table~\ref{tb:40coa_model_2} belongs to the prediction rather than the inference framework \citep{gareth2013introduction} and enables us to identify a prompt that maximizes the success rate of \textbf{gpt-4.1} in obtaining the optimal 16-run 9-factor design. To this end, Table \ref{tb:example_sequences} shows the five tests with the highest predicted success rates among all $(5!)(2)=240$ possible tests, together with their standard errors (SE) adjusted for overdispersion, and actual success rates in a follow-up confirmation experiment with 60 independent runs of \textbf{gpt-4.1} for each test. Among the five tests, three were not in our 80-run OF design.

Overall, the predicted success rates of our model are close to the actual ones from the confirmation experiment. Specifically, the sequence 13524 with INSTRUCTION has a success rate of 98.3\%. For comparison, the base prompt (with sequence 12345 and INSTRUCTION) has a success rate of only 11.7\%. Therefore, OofA experiments enabled us to improve the success rate of generating the 16-run 9-factor optimal design using \textbf{gpt-4.1} by rearranging the prompt elements.

\begin{table}%[ht]
\centering
\caption{Predicted top sequences and success rates obtained with \textbf{gpt-4.1}.}
\label{tb:example_sequences}
\begin{tabular}{ll cccc l}
\toprule
&           &          & \multicolumn{2}{c}{Success rate} & \\ \cmidrule{4-5}
& Sequence & $w_{2}$ & Actual & Predicted & SE & In dataset?\\
\midrule
   & 15234    &  $-1$ & 0.900   & 0.956 & {0.019} & No \\
 & 15324    &  $-1$ & {0.800}  & 0.956 & {0.019} & Yes (test 21)\\
   & 13524    &  $-1$ & 0.983 & 0.951 & {0.023}& No\\
   & 31524    &  $-1$ & 0.867 & 0.949 & {0.025} & No\\
 & 52134    &  $-1$ & {0.933} & 0.927 & {0.033} & Yes (test 35)\\
\bottomrule
\end{tabular}
\end{table}

\iffalse %
\revb{(For ourselves; no need to include in the paper)
Here is an explanation of the large difference on actual and predicted outcome using sequence 15324.
We need to consider the prediction interval (PI) for a future single validation with adjustment for multiple comparisons   (cv=$z_{.05/10}$=2.576; cv=$z_{.01/10}$=3.09).
The 95\% PI lower bound = $0.956-2.576*\sqrt{0.019^2+0.8*0.2/60}=0.814$;
The 99\% PI lower bound = $0.956-3.09*\sqrt{0.019^2+0.8*0.2/60}=0.786$.
%If use p=0.7, 95\% PI lower bound = $0.956-2.576*\sqrt{0.019^2+0.7*0.3/60}=0.796$; that is, even p=0.7 is consistent with the data. If use p=0.5, 95\% PI lower bound = 0.783. 
}
\fi %

\section{Concluding remarks} \label{sec:conclusion}

In this paper, we showed that OofA experiments can be used as a prompt engineering method to assess and deal with order dependency in pretrained transformer-based LLMs. Specifically, we showed that order dependency is present in \textbf{gpt-4.1} and \textbf{gemini-2.5-flash} when constructing two-level fractional factorial designs with 16 runs. We discuss three OofA experiments to quantify and overcome this phenomenon at different degrees. In the first experiment, we estimated the effects of orderings and versions of specific phrases within the instruction submitted to \textbf{gpt-4.1} for constructing designs with seven to 12 factors. A remarkable finding was the significant negative effect of specifying the desired number of runs before the desired number of factors in the instruction for many of these tasks. In the second experiment, we studied the ordering effects of our prompt elements in Section~\ref{sec:promptelements} in the performance of \textbf{gpt-4.1}. In the third experiment, discussed in Supplementary Section~S3, we did the same but using \textbf{gemini-2.5-flash}. Using an ordering factorial design and logistic PWO models, we optimized the success rates of these LLMs in constructing designs with nine and seven factors. Compared to our base prompt, the gains in success rates by \textbf{gpt-4.1} and \textbf{gemini-2.5-flash} were 86.6\% and 65\%, respectively, in these tasks. 

Generative artificial intelligence is evolving rapidly, with new versions of LLMs released almost every month. In fact, when we were writing this article from January to June 2026, OpenAI and Google released new versions of \textbf{gpt-4.1} and \textbf{gemini-2.5-flash}, respectively. However, it turns out that some of these newer LLMs also exhibit order dependency on our tasks. To show this, Supplementary Section~S4 provides additional OofA experiments involving \textbf{gpt-5.4} and \textbf{gemini-3-flash-preview} both with and without their reasoning feature enabled. In these experiments, we studied the success rates of these LLMs in constructing the 16-run 9-factor optimal design. To demonstrate order dependency, we used Prompt~\ref{ptp:basic} and changed the position of the (original) INSTRUCTION element. That is, we used five sequences obtained by placing INSTRUCTION at each of the five positions in Prompt~\ref{ptp:basic}, while keeping the relative ordering of all other prompt elements fixed. Supplementary Section~S4 shows that the success rates of \textbf{gpt-5.4} and \textbf{gemini-3-flash-preview} were lower than 18\% for the five sequences when their reasoning feature was not enabled. When it was enabled, the success rates ranged from 85\% to 97.5\% and from 42.5\% to 65\%, respectively. Therefore, these LLMs are also candidates for our prompt engineering method involving OofA experiments, but we leave this application for future research.

Our application of OofA experiments for prompt engineering suggests several directions for future research. For example, our first experiment motivates the development of OofA designs and models for a \textit{nested} order structure. Recall that we considered two versions of the design dimension component, namely ``$m$ factors and $n$ runs'' and its rephrased version ``$n$ runs and $m$ factors.'' Rather than treating these as two levels of a single component, an alternative is to regard the phrases ``$n$ runs'' and ``$m$ factors'' as two sub-components within the design dimension component. This would allow us to study both ordering effect of the sub-components within the main component---as well as its position effect relative to the other main components---on the performance of an LLM. This idea can be extended by introducing additional sub-components, such as ``$l$ levels'', where $l$ could be two or three. The study of sub-components and main components in this nested structure thus calls for novel cost-efficient OofA designs and extensions to the PWO model or the component position model \citep{yang2021component}.   

Another avenue for future research is to construct custom OofA designs to estimate the logistic PWO models in Section~\ref{sec:logisticPWOmodel} using an optimal design approach \citep{berger2009introduction}. To this end, we could start by considering locally optimal designs which are constructed assuming good initial values for the coefficients in the model under study. Such initial values can be set according to, for example, our numerical results. To this end, we could consider extending the metaheuristic algorithms of \cite{stokes2024metaheuristic} developed to obtain optimal designs for the standard PWO and component position models. 

Finally, although we restricted ourselves to constructing two-level fractional factorial designs using LLMs, we could employ OofA experiments to study order dependency on other types of tasks. For instance, we could study the order of answer options in multiple-choice benchmarks such as CommonsenseQA \citep{talmor2019commonsenseqa} or MMLU \citep{hendrycks2020measuring}, and the order of few-shot demonstrations in GSM8K \citep{cobbe2021training}. Compared to benchmark prompt engineering methods, an advantage of our method is the ability to estimate the effect of the positions of the prompt elements. This allows us to explain the performance of an LLM on a task in terms of the prompt configuration used. The application of OofA experiments to other tasks is another future research direction.

\begin{center}
{\large\bf SUPPLEMENTARY MATERIAL}
\end{center}

\noindent \textbf{Programs.zip}. Zip file containing Python and R code to reproduce all our OofA experiments.

\noindent \textbf{Supplementary sections.pdf}. Pdf file with sections containing additional details on our OofA experiments.

\begin{center}
{\large\bf ACKNOWLEDGMENTS}
\end{center}
The authors acknowledge that ChatGPT and Gemini were used for constructing two-level fractional factorial designs and for language assistance in some sentences through Writefull (\url{https://www.writefull.com/}). They were not used to generate ideas or classify literature. 
\if0\blind
The research of Vazquez is supported by the Challenge-Based Research Fund of Tecnologico de Monterrey under the project CI-EIC-HLT-D-58.  
\fi
\if1\blind
$ $ \\ $ $ \\ 
\fi

\begin{center}
{\large\bf DISCLOSURE OF INTEREST}
\end{center}
The authors report that there are no competing interests to declare.

\begin{center}
{\large\bf DECLARATION OF FUNDING}
\end{center}
No funding was received.

\begin{center}
{\large\bf DATA AVAILABILITY STATEMENT}
\end{center}
All data and code related to this paper are in the supplementary material.

%---------------------------------------------------------------
% References
%---------------------------------------------------------------
%\clearpage
\bibliographystyle{apalike} % Plain referencing style
\bibliography{LLMdoe} % Use the example bibliography file sample.bib

%%%%%%%%%%%%%%%%%% supplementary sections

\clearpage
\setcounter{page}{1}
\setcounter{section}{0}
\def\thesection{S\arabic{section}}
\setcounter{equation}{0}
\def\theequation{S\arabic{equation}}
\setcounter{table}{0}
\def\thetable{S\arabic{table}}

\begin{center}
\textbf{\Large Supplementary Sections} \\ \bigskip

{Duoduo Danny Ying, Alan R. Vazquez, and Hongquan Xu} \\ \medskip

%{\em University of California, Los Angeles} 
\end{center}
%\appendix

\section{Significance of the SHOT Element}

Table \ref{tb:example_seq} shows the success rates among 60 replicates of six tests used to construct the optimal 16-run 9-factor design using \textbf{gpt-4.1}. These tests involve the original INSTRUCTION but different sequences of the prompt elements in Section~3.2 in the main text. The first five sequences were chosen so that the SHOT element appears in each of the five possible positions, whereas the last sequence omits this element entirely. These tests thus allow us to compare the contribution of SHOT in searching for the optimal 16-run 9-factor design. 

\begin{table}[h!]
\centering
\caption{Success rates of selected tests for constructing the 16-run 9-factor optimal design with \textbf{gpt-4.1}.}
\label{tb:example_seq}
\begin{tabular}{cccc}
\toprule
& Sequence & $w_{2}$ & Success rate\\
\midrule
 & 51234    &  {$-1$}& 0.817  \\
   & 15234    & {$-1$}& 0.900 \\
   & 12534    &  {$-1$}& 0.517 \\
   & 12354    &  {$-1$}& 0.300 \\
   & 12345    &  {$-1$}& 0.117 \\
   & 1234    &  {$-1$}& 0.083 \\
\bottomrule
\end{tabular}
\end{table}

Table \ref{tb:example_seq} shows a substantial difference in success rates between the first five sequences and the last sequence, indicating the importance of including SHOT. Moreover, the variation in success rates among the first five sequences shows that its position within the prompt also has a significant effect on the performance of \textbf{gpt-4.1}. This again supports the conclusion that this LLM exhibits order dependency in this task.

\section{Sequential Analysis of Deviance}

Table \ref{tb:anova3} presents the sequential analysis of deviance for the logistic PWO models fitted to the data in Table~1 in the main text. Across the six design construction tasks, the fitted model explains the largest proportion of deviance for the 9-factor design task (76\%) and the smallest proportion for the 11-factor design task (29\%). The contribution of the component ordering effects also varies across tasks, accounting for as much as 45\% of the deviance for the 9-factor design task, but only 19\% for the 11-factor design task. 

Among the version effects, that of design dimensions ($w_2$), which specifies the ordering of the numbers of factors and runs, is the most influential for  four out of the six tasks, explaining between 12\% of the deviance for the 8-factor design task and 28\% for the 12-factor design task. 

Overall, these results demonstrate that both the prompt component orderings and versions have substantial effects on the success rates of optimal design construction.

\begin{table}[h!]
\renewcommand{\tabcolsep}{4pt}
\begin{center}
\caption{Sequential Analysis of Deviance for Table~1 in the main text. }\label{tb:anova3}
\begin{tabular}{ll rrrrrr}
\toprule
& & \multicolumn{6}{c}{Number of Factors} \\ \cline{3-8}
   Source & DF         & 7          & 8         & 9         & 10        & 11        & 12 \\ \hline
%Sequence & 5 & 122.7(29\%) & 330.7(36\%) & 88.1(45\%) & 127.4(37\%) & 31.1(19\%) & 147.7(30\%)  \\
%Version & 7 & 111.5(26\%) & 144.8(16\%) & 63.2(32\%) & 101.2(30\%) & 20.8(13\%) & 170.3(35\%)  \\
%Residual & 35 & 191.8(45\%) & 438.9(48\%) & 43.2(22\%) & 112.8(33\%) & 108.9(68\%) & 174.3(35\%)  \\
%\midrule
Sequence & 5 & 122.7 (29\%) & 330.7 (36\%) & 88.1 (45\%) & 127.4 (37\%) & 31.1 (19\%) & 147.7 (30\%)  \\
$w_1$ & 1 & 16.2 (4\%) & 9.8 (1\%) & 6.9 (4\%) & 2.9 (1\%) & 3.9 (2\%) & 20.1 (4\%)  \\
$w_2$ & 1 & 4.1 (1\%) & 111 (12\%) & 41.8 (21\%) & 63.8 (19\%) & 7.6 (5\%) & 139.5 (28\%)  \\
$w_3$ & 1 & 19.8 (5\%) & 3.5 (0\%) & 11.3 (6\%) & 16.4 (5\%) & 4.4 (3\%) & 0.1 (0\%)  \\
Residual & 39 & 263.1 (62\%) & 459.3 (50\%) & 46.4 (24\%) & 130.9 (38\%) & 113.9 (71\%) & 184.9 (38\%)  \\
\bottomrule
\end{tabular}
\end{center}
Note: DF stands for degrees of freedom.
\end{table}

%%\section{Details on the Third OofA Experiment}
\section{Third Experiment with gemini-2.5-flash}

In this section, we discuss the third OofA experiment, which aims to optimize the performance of \textbf{gemini-2.5-flash} in constructing optimal two-level designs. As a first step, we identify a task that exhibits order dependency and is thus suitable for optimization using our approach. To this end, we consider the 16-run design construction tasks with seven to 12 factors, and evaluate specific sequences of the five prompt elements in Section~3.2 in the main text. Specifically, the five sequences are constructed by placing FORMAT at each of the five possible positions, while keeping the relative ordering of the remaining components fixed. All sequences involve the original INSTRUCTION element.

Table~\ref{tb:gemini_2_5_task} shows the success rates in 20 replicates obtained from these sequences using \textbf{gemini-2.5-flash}. The table shows significant differences among the five sequences when the task has seven and eight factors, indicating that the order of the prompt elements has a noticeable impact on the success rate. In contrast, for tasks with nine or more factors, the success rates are virtually zero, leaving little variation between the sequences. For illustration, we therefore focus on the 16-run 7-factor design construction task, which has clear order effects and the potential for a successful optimization using OofA experiments. 

\begin{table}[h!]
\centering
\caption{Success rates of \textbf{gemini-2.5-flash} for constructing 16-run optimal designs with seven to 12 factors.} \label{tb:gemini_2_5_task}
\begin{tabular}{llllllll}
\toprule
&  & \multicolumn{6}{c}{Number of Factors} \\ \cline{3-8}

Sequence   & $w_2$ & 7   & 8   & 9   & 10       & 11 & 12 \\
\midrule
41235  & $-1$  & 0.4 & 0.25  & 0.1  & 0.05  & 0  & 0.05  \\
14235  & $-1$  & 0.15  & 0.35 & 0 & 0.05 & 0 & 0  \\
12435  & $-1$  & 0.55  & 0.4 & 0.05 & 0.1 & 0 & 0.05  \\
12345  & $-1$  & 0.35 & 0.45 & 0.05 & 0 & 0 & 0.05  \\
12354  & $-1$  & 0.8 & 0.5 & 0 & 0 & 0 & 0.05  \\
\bottomrule
\end{tabular}
\end{table}

\begin{figure}[thbp]
    \centering
    \includegraphics{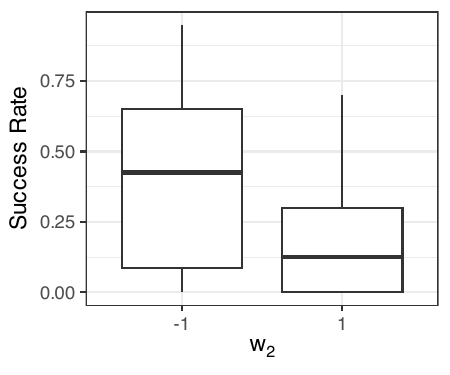}
    \caption{Success rates of \textbf{gemini-2.5-flash} for constructing 16-run 7-factor optimal designs across two versions of INSTRUCTION.}
    \label{fig:FactorRun-Gemini}
\end{figure}

To identify the optimal prompt for the 16-run 7-factor design construction task, we conducted an OofA experiment according to the 80-run OF design in Table~3 in the main text. Table~\ref{tb:40coadesign_gemini} shows the success rates obtained from each of the 80 tests using 20 independent executions of \textbf{gemini-2.5-flash}. Figure~\ref{fig:FactorRun-Gemini} shows the distribution of the responses from Table \ref{tb:40coadesign_gemini} under the two levels of component 2 (INSTRUCTION). When $w_2 = -1$ (corresponding to the original INSTRUCTION), \textbf{gemini-2.5-flash} has an average success rate of 39.4\%, compared with only 18.6\% when $w_2 = 1$ (corresponding to INSTRUCTION$_1$). So, specifying the desired number of factors before the desired number of runs is more effective than the reverse ordering. The observed pattern is consistent with that for \textbf{gpt-4.1}, indicating that \textbf{gemini-2.5-flash} is also sensitive to changes in the elements of a prompt.

\begin{table}
\centering
\small
\caption{Ordering factorial design and success rates of \textbf{gemini-2.5-flash} for constructing the optimal 16-run 7-factor design.}\label{tb:40coadesign_gemini}
\begin{tabular}{ccrc|ccrc|ccrc}
\toprule
Test & Sequence & $w_2$ & Rate & Test & Sequence & $w_2$ & Rate & Test & Sequence & $w_2$ & Rate\\ \hline
1 & 14325 & $-1$ & 0.55 & 28 & 31254 & $-1$ & 0.70 & 55 & 54132 & 1 & 0.00\\
2 & 25431 & $-1$ & 0.60 & 29 & 42315 & $-1$ & 0.65 & 56 & 13452 & 1 & 0.30\\
3 & 31542 & $-1$ & 0.55 & 30 & 53421 & $-1$ & 0.05 & 57 & 24513 & 1 & 0.55\\
4 & 42153 & $-1$ & 0.35 & 31 & 13245 & $-1$ & 0.80 & 58 & 35124 & 1 & 0.00\\
5 & 53214 & $-1$ & 0.00 & 32 & 24351 & $-1$ & 0.75 & 59 & 41235 & 1 & 0.45\\
6 & 12534 & $-1$ & 0.65 & 33 & 35412 & $-1$ & 0.10 & 60 & 52341 & 1 & 0.00\\
7 & 23145 & $-1$ & 0.65 & 34 & 41523 & $-1$ & 0.00 & 61 & 15324 & 1 & 0.00\\
8 & 34251 & $-1$ & 0.60 & 35 & 52134 & $-1$ & 0.00 & 62 & 21435 & 1 & 0.30\\
9 & 45312 & $-1$ & 0.25 & 36 & 12453 & $-1$ & 0.50 & 63 & 32541 & 1 & 0.10\\
10 & 51423 & $-1$ & 0.00 & 37 & 23514 & $-1$ & 0.60 & 64 & 43152 & 1 & 0.00\\
11 & 15243 & $-1$ & 0.65 & 38 & 34125 & $-1$ & 0.10 & 65 & 54213 & 1 & 0.10\\
12 & 21354 & $-1$ & 0.90 & 39 & 45231 & $-1$ & 0.15 & 66 & 14532 & 1 & 0.05\\
13 & 32415 & $-1$ & 0.85 & 40 & 51342 & $-1$ & 0.00 & 67 & 25143 & 1 & 0.40\\
14 & 43521 & $-1$ & 0.00 & 41 & 14325 & 1 & 0.10 & 68 & 31254 & 1 & 0.20\\
15 & 54132 & $-1$ & 0.00 & 42 & 25431 & 1 & 0.70 & 69 & 42315 & 1 & 0.15\\
16 & 13452 & $-1$ & 0.60 & 43 & 31542 & 1 & 0.20 & 70 & 53421 & 1 & 0.05\\
17 & 24513 & $-1$ & 0.60 & 44 & 42153 & 1 & 0.15 & 71 & 13245 & 1 & 0.30\\
18 & 35124 & $-1$ & 0.00 & 45 & 53214 & 1 & 0.00 & 72 & 24351 & 1 & 0.25\\
19 & 41235 & $-1$ & 0.30 & 46 & 12534 & 1 & 0.15 & 73 & 35412 & 1 & 0.05\\
20 & 52341 & $-1$ & 0.00 & 47 & 23145 & 1 & 0.60 & 74 & 41523 & 1 & 0.05\\
21 & 15324 & $-1$ & 0.10 & 48 & 34251 & 1 & 0.00 & 75 & 52134 & 1 & 0.00\\
22 & 21435 & $-1$ & 0.85 & 49 & 45312 & 1 & 0.35 & 76 & 12453 & 1 & 0.30\\
23 & 32541 & $-1$ & 0.95 & 50 & 51423 & 1 & 0.00 & 77 & 23514 & 1 & 0.40\\
24 & 43152 & $-1$ & 0.15 & 51 & 15243 & 1 & 0.50 & 78 & 34125 & 1 & 0.20\\
25 & 54213 & $-1$ & 0.15 & 52 & 21354 & 1 & 0.40 & 79 & 45231 & 1 & 0.05\\
26 & 14532 & $-1$ & 0.25 & 53 & 32415 & 1 & 0.00 & 80 & 51342 & 1 & 0.00\\
27 & 25143 & $-1$ & 0.80 & 54 & 43521 & 1 & 0.05 &  &  &  & \\
\bottomrule
\end{tabular}
\end{table}

To continue with our statistical analysis, we fit the logistic PWO model in Equation~(5) in the main text to the data in Table~\ref{tb:40coadesign_gemini}. Table \ref{tb:40coa_gemini} shows the estimated coefficients of this model. Consistent with the findings for \textbf{gpt-4.1}, the ordering of the phrases specifying the number of factors and runs is the most influential effect. The negative estimate of the coefficient for $w_{2}$ indicates that placing the number of factors before the number of runs increases the success rate in generating the 16-run 7-factor optimal design. 

\begin{table}%[ht]
\centering
\caption{Estimated coefficients and associated statistics of a logistic PWO model for the 16-run 7-factor design construction task with \textbf{gemini-2.5-flash}.}
\label{tb:40coa_gemini}
\begin{tabular}{lrrrl}
\hline
            & Estimate  & Standard Error & z value & Pr($>|z|$) \\
\hline
(Intercept)       & $-$1.210   & 0.159     & $-$7.602    & 2.91e-14 \\
$z_{12}$    & $-$0.101    & 0.212     & $-$0.475    & 0.635  \\
$z_{13}$    & $-$0.027    & 0.210      & $-$0.126    & 0.900   \\
$z_{14}$    & 0.149     & 0.213     &  0.700      & 0.484   \\
$z_{15}$    & 0.237     & 0.210      &  1.130     & 0.258 \\
$z_{23}$    & 0.048     & 0.212     &  0.225    & 0.822  \\
$z_{24}$    & 0.373     & 0.210      &  1.781   & 0.075  \\
$z_{25}$    & 0.687     & 0.212     &  3.248    & 0.001  \\
$z_{34}$    & $-$0.338    & 0.211     & $-$1.603   & 0.109  \\
$z_{35}$    & 0.273     & 0.215     &  1.267    & 0.205  \\
$z_{45}$    & $-$0.112    & 0.209     & $-$0.538    & 0.591  \\
$w_{2}$   & $-$0.643     & 0.146     & $-$4.401  & 1.08e-5  \\
\hline
\multicolumn{5}{l}{Null deviance: 712.75 on 79 degrees of freedom} \\
\multicolumn{5}{l}{Residual deviance: 315.54  on 68 degrees of freedom} \\
%Null deviance     & 712.75 \\
%Residual deviance & 315.54 \\
$R^2_{McF}$       & 0.557 \\
$\hat\phi$        & 4.966 \\
\hline
\end{tabular}
\end{table}

Among the component ordering effects, $z_{25}$ is the most significant as its p-value is approximately 0.001. At a level of $\alpha = 0.1$, $z_{24}$ is also statistically significant. The positive estimates for both $z_{25}$ and $z_{24}$ indicate that placing component 2 (INSTRUCTION) before component 4 (FORMAT) and 5 (SHOT) is associated with a higher success rate. This finding differs from the results for \textbf{gpt-4.1}, where the preferred ordering places SHOT before INSTRUCTION. The difference suggests that effective element orderings depend on the underlying LLMs.%%, something that was mentioned by \cite{lu2022fantastically}. 

Although the logistic PWO model achieves a $R^2_{McF}$ of 0.557, indicating a reasonable fit, its explanatory power can be improved by incorporating two-way interactions among the pseudo factors and $w_2$. Because the full interaction model contains a large number of terms, we use a step-wise regression with the Bayesian Information Criterion to obtain a reduced model. The resulting model is shown in Table \ref{tb:40coa_gemini_BIC}. Relative to the logistic  PWO model in Table \ref{tb:40coa_gemini}, the final reduced model includes eight additional terms and an $R^2_{McF}$ of 0.789. Furthermore, the estimated dispersion parameter $\hat{\phi}$ decreases from 4.966 to 2.33. Therefore, incorporating two-way interactions yields a considerably better fit than a model containing main effects only.

Table~\ref{tb:40coa_gemini_BIC}  reports the estimated coefficients and associated statistics for the reduced model. The variable $w_2$ remains the most significant, followed by $z_{25}$, $z_{34}z_{35}$ and $z_{15}z_{25}$. All terms except $z_{12}$ appear as a main effect and within at least one interaction term. Comparing these results with those for \textbf{gpt-4.1} in Table~5 in the main text, we have that $w_2$ is still the dominant effect in both models, whereas the remaining significant terms differ substantially. 

\begin{table}%[ht]
\centering
\caption{Estimated coefficients and associated statistics of a reduced logistic PWO model with interaction effects for 16-run 7-factor design construction task with \textbf{gemini-2.5-flash}.} \label{tb:40coa_gemini_BIC}
\begin{tabular}{lrrrl}
\hline
            & Estimate  & Standard Error & z value & Pr($>|z|$) \\
\hline
(Intercept) & $-$1.467    & 0.190     & $-$7.721    & 1.15e-14 \\
$z_{13}$    & $-$0.085   & 0.163   &  $-$0.521    & 0.603  \\
$z_{14}$    & 0.020     & 0.164   & 0.122   & 0.903   \\
$z_{15}$    & 0.502    & 0.156   &  3.221   & 0.001 \\
$z_{23}$    & 0.227    & 0.163   &  1.391   & 0.164  \\
$z_{24}$    & 0.146    & 0.164   &  0.893   & 0.372 \\
$z_{25}$    & 0.865    & 0.168   & 5.153    & 2.56e-7 \\
$z_{34}$    & $-$0.349   & 0.170   &  $-$2.054   & 0.04  \\
$z_{35}$    & 0.340     & 0.183   &  1.858    & 0.063  \\
$z_{45}$    & $-$0.179   & 0.150   & $-$1.190    & 0.234  \\
$w_{2}$   & $-$0.64     & 0.111   & $-$5.757  & 8.57e-9  \\
$z_{15}z_{25}$   & $-$0.506  & 0.129   & $-$3.920  & 8.84e-5  \\
$z_{14}z_{15}$   & 0.399   & 0.134   & 2.982  & 0.003  \\
$z_{34}z_{35}$   & 0.576   & 0.147   & 3.924  & 8.72e-5  \\
$z_{13}z_{35}$   & 0.411     & 0.144     & 2.848  & 0.004  \\
$z_{13}z_{45}$   & $-$0.255     & 0.121     & $-$2.097  & 0.040  \\
$z_{23}z_{25}$   & 0.193     & 0.130     & 1.486  & 0.137 \\
$z_{35}w_{2}$   & $-$0.241     & 0.115     & $-$2.098  & 0.036  \\
$z_{24}w_{2}$   & $-$0.259     & 0.114     & $-$2.279  & 0.023  \\
$z_{23}w_{2}$   & 0.176     & 0.123     & 1.425  & 0.154 \\
\hline
\multicolumn{5}{l}{Null deviance: 712.75 on 79 degrees of freedom}\\
\multicolumn{5}{l}{Residual deviance: 150.6 on 60 degrees of freedom}
 \\
$R^2_{McF}$       & 0.789 \\
$\hat\phi$        & 2.33 \\
\hline
\end{tabular}
\end{table}

We then used the reduced logistic PWO model with interactions to predict the success rates of all 240 possible prompt sequences. We selected the five tests with the highest predicted rates for confirmation experiments. Table~\ref{tb:example_sequences_gemini} shows the results from these experiments. In this table, the success rate is obtained using 60 independent executions of \textbf{gemini-2.5-flash}. The table reports the predicted success rate, the corresponding standard error after adjusting for overdispersion, and the observed success rate from each confirmation test. 

Overall, Table~\ref{tb:example_sequences_gemini} shows that the predicted and observed rates agree for each test, suggesting that the fitted model provides accurate predictions. Moreover, the average success rate increases from approximately 35\% under the original sequence in Table~\ref{tb:gemini_2_5_task} to 100\% for the selected sequences. These substantial improvements demonstrate that OofA experiments can effectively generate high-performing prompts for \textbf{gemini-2.5-flash}. %%In fact, it can be used to optimize prompt ordering for any LLM that demonstrates the order dependency problem.

\begin{table}[h!]
\centering
\caption{Predicted top sequences and success rates for constructing the 16-run 7-factor optimal design {with \textbf{gemini-2.5-flash}}.}
\label{tb:example_sequences_gemini}
\begin{tabular}{ll cccc l}
\toprule
&           &          & \multicolumn{2}{c}{Success rate} & \\ \cmidrule{4-5}
& Sequence & $w_{2}$ & Actual & Predicted & SE & In dataset?\\
\midrule
 & 21354    &  {$-1$} & 1.000   & 0.935 & 0.033 & Yes (test 12)\\
   & 12354    & {$-1$} & 0.880  & 0.935 & 0.033 & No\\
   & 13254    &  {$-1$} & 0.880 & 0.899 & 0.046 & No\\
   & 24531    &  {$-1$} & 0.820 & 0.879 & 0.057 & No\\
   & 13524    &  {$-1$} & 0.700 & 0.864 & 0.069 & No\\
\bottomrule
\end{tabular}
\end{table}

\section{Experiments on Other LLMs}

Here, we demonstrate that order dependency persists in more recent GPT and Gemini models, including \textbf{gpt-5.4} and \textbf{gemini-3-flash-preview}. To illustrate this phenomenon, we conducted a small experiment to construct the 16-run 9-factor optimal design using these LLMs. In this experiment, we considered the five elements in Section~3.2 in the main text. We considered five sequences obtained by placing component 2 (INSTRUCTION) at each of the five possible positions while keeping the relative ordering of all other prompt elements fixed. For each sequence, we constructed the corresponding prompt and submitted it 40 times to each LLM. Table~\ref{tb:result_gpt_gemini} shows the success rates obtained by \textbf{gpt-5.4} and \textbf{gemini-3-flash-preview} using both \textit{no} and \textit{low} thinking settings. 

Table~\ref{tb:result_gpt_gemini} shows substantial differences in performance across the five sequences for both models under both settings. For instance, the success rate of \textbf{gpt-5.4} ranges from 85\% to 97.5\% when using low thinking. For \textbf{gemini-3-flash-preview}, the success rate ranges from 42.5\% to 65\% under a similar setup. Therefore, these newer-generation LLMs still exhibit order dependency, thereby necessitating OofA experiments to optimize their performance. 

\begin{table}[h!]
\centering
\caption{Success rates for constructing 16-run 9-factor optimal designs using no or low thinking with \textbf{gpt-5.4} and \textbf{gemini-3-flash-preview}.}
\label{tb:result_gpt_gemini}
\begin{tabular}{ll cc c cc}
\toprule
      & &   \multicolumn{2}{c}{\textbf{gpt-5.4}} & & \multicolumn{2}{c}{\textbf{gemini-3-flash-preview}} \\
      \cmidrule{3-4} \cmidrule{6-7}
Sequence   & $w_2$ & No Thinking   & Low Thinking   & & No Thinking   & Low Thinking \\
\midrule
21345  & $-1$  & 0.000      & 0.975 & & 0.025 & 0.550   \\
12345  & $-1$  & 0.150   & 0.975 & & 0.175 & 0.425  \\
13245  & $-1$  & 0.025  & 0.975 & & 0.075 & 0.550 \\
13425  & $-1$  & 0.025  & 0.950 & & 0.025  & 0.625 \\
13452  & $-1$  & 0.175  & 0.850 & & 0.125  & 0.650 \\
\bottomrule
\end{tabular}
\end{table}

\end{document}